\documentclass[nolinenumbers]{aastex631}

\shorttitle{Possible TeV $\gamma$-ray binary origin of HESS J1828-099}
\shortauthors{De Sarkar et al.}
\graphicspath{{./}{figures/}}
\usepackage{xcolor}
\usepackage{graphicx}

\begin{document}

\title{Possible TeV $\gamma$-ray binary origin of HESS J1828-099}

\author[0000-0001-6047-6746]{Agnibha De Sarkar}
\affiliation{Astronomy $\&$ Astrophysics group, Raman Research Institute \\
C. V. Raman Avenue, 5th Cross Road, Sadashivanagar, Bengaluru 560080, Karnataka, India}
\email{agnibha@rri.res.in}

\author[0000-0001-9829-7727]{Nirupam Roy}
\affiliation{Dept. of Physics, Indian Institute of Science \\
CV Raman Road, Bengaluru 560012, Karnataka, India}

\author[0000-0002-5481-5040]{Pratik Majumdar}
\affiliation{Saha Institute of Nuclear Physics\\
A CI of Homi Bhabha National Institute, Kolkata 700064, West Bengal, India}

\author[0000-0002-1188-7503]{Nayantara Gupta}
\affiliation{Astronomy $\&$ Astrophysics group, Raman Research Institute \\
C. V. Raman Avenue, 5th Cross Road, Sadashivanagar, Bengaluru 560080, Karnataka, India}

\author[0000-0003-4468-761X]{Andreas Brunthaler}
\affiliation{Max Planck Institut f\"ur Radioastronomie\\
Auf dem H\"ugel 69, D-53121 Bonn, Germany}

\author[0000-0001-6459-0669]{Karl M. Menten}
\affiliation{Max Planck Institut f\"ur Radioastronomie\\
Auf dem H\"ugel 69, D-53121 Bonn, Germany}

\author[0000-0001-6010-6200]{Sergio A. Dzib}
\affiliation{Max Planck Institut f\"ur Radioastronomie\\
Auf dem H\"ugel 69, D-53121 Bonn, Germany}

\author[0000-0001-9536-7494]{Sac Nict\'e X. Medina}
\affiliation{Max Planck Institut f\"ur Radioastronomie\\
Auf dem H\"ugel 69, D-53121 Bonn, Germany}

\author[0000-0003-4516-3981]{Friedrich Wyrowski}
\affiliation{Max Planck Institut f\"ur Radioastronomie\\
Auf dem H\"ugel 69, D-53121 Bonn, Germany}

\begin{abstract}

The High Energy Stereoscopic System (H.E.S.S.) observatory has carried a deep survey of the Galactic plane, in the course of which the existence of a significant number of ($\sim$ 78) TeV $\gamma$-ray sources was confirmed, many of which remain unidentified. HESS J1828-099 is a point-like (Gaussian stand. dev. $<$ 0.07$^{\circ}$) unidentified source among the 17 confirmed point-like sources in the H.E.S.S. Galactic Plane Survey (HGPS) catalog. This source is also unique because it does not seem to have any apparent association with any object detected at other wavelengths. We investigate the nature and association of HESS J1828-099 with multi-wavelength observational data. A high mass X-Ray binary (HMXB) $\textendash$ comprising of pulsar XTE J1829-098 and a companion Be star $\textendash$ has been observed earlier in the X-ray and infrared bands, 14$'$ away from HESS J1828-099. With 12 years of $\textit{Fermi}$-LAT  $\gamma$-ray data, we explore the possibility of 4FGL J1830.2-1005 being the GeV counterpart of HESS J1828-099. Within the RXTE confidence region, a steep spectrum ($\alpha_{radio}$ = - 0.746 $\pm$ 0.284), plausible counterpart is detected in data from existing radio frequency surveys. In this letter, we probe for the first time using multi-wavelength data, whether HESS J1828-099, 4FGL J1830.2-1005 and the HMXB system have a common origin. Our study indicates that HESS J1828-099 might be a TeV high mass $\gamma$-ray binary source.   

\end{abstract}


\section{Introduction} \label{sec:intro}

High mass $\gamma$-ray binaries (HMGBs) belong to a special class of HMXBs, which mainly emit in $\gamma$-ray energies \citep{Dubus_13}. Such objects comprise of compact objects such as a neutron star or a black hole and an O or Be type star as the companion. The $\gamma$-ray emission in such binaries is usually assumed to be powered by wind-driven shocks \citep{Dubus_13}. The compact object in the HMGBs, usually a rotation-powered pulsar, dissipates its rotational energy, by energizing pair plasma, which interacts with wind from the companion star \citep{Maraschi_81, Dubus_06, Huber_21}. In a close orbit system, a wind collision region is created due to this interaction, which in turn terminates the pulsar and stellar winds by a shock \citep{Bogavalov_08, Bosch_12, Huber_21}. Particles can be accelerated to ultra-relativistic energies at these shock sites due to diffusive shock acceleration, later producing observed emission through various radiative processes \citep{Sironi_11, Huber_21}. Another favored emission  scenario can occur if the massive companion star is Be star with a disk. In this scenario, the primary interaction happens as the pulsar crosses the circumstellar disk of the Be star, as in the cases of PSR B1259-63 \citep{aharonianetal05} and PSR J2032+4127 \citep{Lyne_15}. The multi-wavelength emission for these two sources differ from other HMGBs, perhaps due to the geometry of the circumstellar decretion disk. For example, in case of PSR B1259-63, the light curve in the radio, X-ray and TeV regimes is typically double-peaked and driven by synchrotron (radio and X-ray emissions) and Inverse Compton (TeV emission) cooling \citep{chernyakova}. The emission in the GeV range is peculiar given that flares that exceed the pulsar spindown luminosity have been observed with \textit{Fermi}-LAT \citep{peri_2010, peri_14, peri_17, peri_21}. Alternatively, the microquasar model, in which interaction primarily occurs in the jets produced by accretion onto a black hole, also cannot be ruled out \citep{romero_03, bosch_04}. Only a handful of objects, which have been detected above 100 MeV, are firmly established as HMGBs. Some of the observed HMGBs are: HESS J0632-057, 1FGL J1018-5658, PSR B1259-63, LS I +61$^\circ$303, LS 5039 \citep{Dubus_15, Li_17}, PSR J2032+4127 \citep{Abeysekara_18, Lyne_15, Ho_17}, a point source in the Large Magellanic Cloud \citep{Corbet_16, HESS_18_1}, 4FGL J1405.1-6119 \citep{Corbet_19} and HESS J1832-093 \citep{Eger_16, Tam_20, Mart_2020}. All of these sources have soft spectra in TeV energies and hard, absorbed spectra in X-ray energies.

HESS J1828-099 is a new Very High Energy (VHE) TeV $\gamma$-ray source that has been detected in the HGPS \citep{HESS_18_2} at the position of R.A. = 18$^h$28$^m$58.72$^s$ and Decl. = -09$^\circ$59$'$33.8$''$ (J2000). This H.E.S.S. source is detected at a confidence level of 8.9$\sigma$ and the size of the source is 0.05$^\circ$ $\pm$ 0.01$^\circ$, making it one of the 17 point like VHE $\gamma$-ray sources found in HGPS catalog. The flux from this TeV source was recorded for a livetime of 46.3 hours and its 0.20 - 61.90 TeV spectrum is well fitted by a power law ($\propto$ E$^{-\Gamma_{TeV}}$) having a photon index of $\Gamma_{TeV}$ = 2.25 $\pm$ 0.12. Its flux is 1.9 $\pm$ 0.3 $\%$ that of the Crab Nebula above 1 TeV and a 1-dimensional Gaussian model was used as a spatial template to fit the extent of this VHE source. This H.E.S.S. source is still unidentified as it does not seem to have any apparent association with any other source at lower energies. Earlier, \cite{neronov_10} claimed that  1FGL J1829.6-1006 (slightly more than 0.25$^\circ$ away from the H.E.S.S. source) could be the GeV counterpart of the TeV source. They also found that the pulsar J1828-1007 is located at 0.1$^\circ$ from the H.E.S.S. source. Moreover, they claimed the spatial separation between the low and high energy emission regions indicates that this source is possibly a pulsar wind nebula (PWN). However, this was not confirmed by the version of the $\textit{Fermi}$-LAT catalog available at that time, i.e. 3FGL catalog \citep{3FGL_15} or 2FHL catalog \citep{2FGL_16}. This pulsar is also absent in the latest 4FGL catalog \citep{abdol_20}.

In this letter, we report our investigations on the origin of the VHE source HESS J1828-099. Analysis of the $\textit{Fermi}$-LAT data revealed a possible GeV counterpart, 4FGL J1830.2-1005, spatially coincident with the H.E.S.S. source. A Galactic X-ray source XTE J1829-098 was also observed by Chandra X-ray observatory, within the 68$\%$ containment radius of 4FGL J1830.2-1005 and 14$'$ away from the centroid of HESS J1828-099 \citep{halpern_07}, making it a very likely lower energy counterpart of both the 4FGL and H.E.S.S. sources, based on its position. Pulsar XTE J1829-098 was observed as a transient source by the Rossi X-ray Timing Explorer (RXTE) observatory, during the scan of the Galactic plane in 2003 July - 2003 August \citep{markwadt_04}. The best-fit pulsar position was found to be R.A. = 18$^h$29$^m$35$^s$ and Decl. = -09$^\circ$51$'$0.00$''$ (J2000), with a 99$\%$ confidence region of approximately elliptical shape, with semimajor axes of 3.8$'$ (RA) and 3$'$ (Decl.) \citep{markwadt_04}. Subsequent X-ray Multi-Mirror Mission (XMM-Newton) observations found the position of this source to be R.A. = 18$^h$29$^m$44.1$^s$ and Decl. = -09$^\circ$51$'$24.1$''$ (J2000), with a 90$\%$ uncertainty radius of 3.2$''$ \citep{halpern_07}. It was discovered in the RXTE data that this pulsar has a rotation period of $\sim$ 7.8 s \citep{markwadt_04}, which was later confirmed by various other observations \citep{halpern_07, shtykovsky_18}. Analyzing XMM-Newton data, a hard power law photon index, $\Gamma_{X}^{XMM}$, of 0.76 $\pm$ 0.13 and a hydrogen column density $N_H$, of (6.0 $\pm$ 0.6) $\times$ 10$^{22}$ cm$^{-2}$ were estimated in the soft X-ray range (2 - 10 keV), both given with their 1$\sigma$ uncertainties \citep{halpern_07}. This suggests that this pulsar is part of a HMXB, as the best-fit value of $N_H$ exceeds the measured Galactic 21 cm HI column density, in the pulsar's direction, of $\sim$ 1.81 $\times$ 10$^{22}$ cm$^{-2}$ \citep{DL_1990}, $\sim$ 1.43 $\times$ 10$^{22}$ cm$^{-2}$ \citep[Leiden/Argentine/Bonn (LAB) survey;][]{kalberla_05}, $\sim$ 1.79 $\times$ 10$^{22}$ cm$^{-2}$ \citep[HI4PI survey;][]{HI4PI}, indicating that some absorption is intrinsic to the binary, either from the wind or circumstellar disk of the companion star. A candidate source, 2.1$''$ away from XMM-Newton location of XTE pulsar, was detected in the analysis of the data obtained by Chandra \citep{halpern_07}. The Chandra location of this source was found to be R.A. = 18$^h$29$^m$43.97$^s$ and Decl. = -09$^\circ$51$'$23.2$''$ (J2000), with a 90$\%$ positional uncertainty of 0.6$''$. Assuming the same best-fit XMM-Newton parameters, the average flux of the source, detected by Chandra in the soft X-Ray range, was found to be consistent with that from the XMM-Newton observations \citep{halpern_07}. A hard, absorbed spectrum estimated from the analysis of archival data obtained by Swift-X-Ray Telescope (XRT) ($\Gamma_X^{Swift}$ = 1.1$^{+0.9}_{-0.8}$, N$_H$ = 10$^{+6}_{-4}$ $\times$ 10$^{22}$ cm$^{-2}$) reinforces this source's identification as a HMXB \citep{sguera_19}. This source has shown frequent outbursts over the years, observed by different observatories. The MAXI gas slit camera (GSC) detected 4 outbursts from this source in 11 years of observation, including one on 2021 April 12 \citep{nakajima_21}. The time intervals between these outbursts matches the proposed orbital period ($\approx$ 246 days) of the binary system \citep{markwadt_04, nakajima_21}. \cite{sguera_19} had checked 15 - 50 keV XTE source light curve on daily timescale from Swift-Burst Alert Telescope (BAT) archive, and found that the duration of the outburst was very likely of the order of 3 - 4 days, which is almost the same order of duration estimated by \cite{markwardt09} ($\sim$ 7 days). In 2018 August, an X-ray outburst from this source triggered a ToO observation with the Nuclear Spectroscopic Telescope Array (NuSTAR), which showed the existence of a cyclotron absorption line at $E_{cyc}$ = 15.05 $\pm$ 0.06 keV, which implies that the magnetic field on the neutron star surface is B $\simeq$ 1.7 $\times$ 10$^{12}$ Gauss \citep{shtykovsky_18}. The detection of the cyclotron absorption line in the X-Ray spectrum of the pulsar confirmed that this pulsar is part of a HMXB.

A star was found in infrared (IR) analysis, within 0.2$''$ of the Chandra localization of XTE J1829-098 \citep{halpern_07}. This bright, infrared counterpart was detected in the Two Micron All Sky Survey (2MASS), but it is not visible in the optical range. The measured IR magnitudes of this companion star are K = 12.7, H = 13.9, I $>$ 21.9 and R $>$ 23.2 \citep{halpern_07}. From the measured magnitude in the H and K bands, the distance of this companion was estimated to be approximately 10 kpc. Assuming this distance, the maximum observed X-Ray luminosity in the 2 - 10 keV range was found to be 2 $\times$ 10$^{36}$ erg\,s$^{-1}$ and minimum luminosity as 3 $\times$ 10$^{32}$ erg\,s$^{-1}$, similar to a wind-driven system or a Be binary transient \citep{halpern_07}. Later observations by \cite{sguera_19} found that reddening free near-infrared (NIR) diagnostic color criterion Q has a value of -0.7, which is very typical of early type OB star, although it can also be a Be star. According to the Corbet diagram \citep{corbet_84, corbet_85, corbet_86}, for a possible orbital period of $\approx$ 246 days, there is a greater likelihood that the donor star is a Be star. Moreover, the absence of an H$\alpha$ emission line in the NIR spectra of the 2MASS counterpart is indicative of the NIR counterpart being a Be star.

Data analysis and the corresponding results are discussed in section \ref{sec:DAR}. In subsection \ref{subsec:XRAY}, we present the results of the analysis of NuSTAR data and report the detection of a sub-dominant, intrabinary shock emission component. Based on this detection and spatial association, we suggest that this HMXB has a common origin with both of the 4FGL and H.E.S.S. sources. In subsection \ref{subsec:FLA}, we present the results of the analysis of $\sim$ 12 years of $\textit{Fermi}$-LAT data. We have also used multiwavelength radio continuum data to identify any radio counterpart of the H.E.S.S. source. In subsection \ref{subsec:RDA}, we discuss the detection of a nearby source in multi radio frequency surveys and investigate this as the likely radio counterpart of the H.E.S.S. source based on its position. In section \ref{SED_modelling}, we present the results of one-zone leptonic modelling to  fit the multi-wavelength spectral energy distribution (SED) and show that the required values of parameters are consistent with those of other established HMGBs \citep{hinton_08}. Finally in section \ref{sec:DAC}, we discuss the results and the caveats of our model. We also suggest the additional observations that are required to completely explain the multiwavelength SED of the system. Finally, we conclude that HESS J1828-099 is possibly a TeV HMGB, based on spatial coincidence and spectral properties.

\section{Data analysis and results} \label{sec:DAR}

\subsection{X-Ray data analysis} \label{subsec:XRAY}

 Although the XTE J1829-098 was confirmed to be a HMXB, the presence of an iron K$\alpha$ emission line, the cyclotron absorption line and the exponential cutoff, as reported in \cite{shtykovsky_18}, point towards the fact that the pulsar is accreting and the dominant X-ray flux seen from this source is due to the accretion. However, in previous analyses of established TeV HMGBs \citep{volkov_21, takahashi_09, an_15}, no spectral lines and/or cutoff or spectral turnover at higher energies were found, indicating,  as in general for TeV HMGBs, that the pulsar usually is not accreting. Also, the best-fit cutoff power law spectral index obtained from NuSTAR data analysis is notably different compared to what is predicted if we assume that the observed X-rays represent synchrotron emission. These factors put the TeV HMGB interpretation of HESS J1828-099 to the question.

To resolve this discrepancy, we tried to find whether or not the pulsar in this case is actively accreting, by comparing the Alfven radius (R$_{Alf}$) with the corotation radius (R$_{co}$). If R$_{Alf}$ $<$ R$_{co}$, then material from the companion star accretes on the pulsar surface, if R$_{Alf}$ $>>$ R$_{co}$, then the stellar material directly interacts with pulsar's rotating magnetosphere and subsequently gets ejected; known as the propeller phase. Finally, if R$_{Alf}$ $\simeq$ R$_{co}$, then these two effects happen simultaneously, and intermittent accretion occurs, which is the intermediate stage of accretor and propeller phases.

The corotation radius (R$_{co}$) is defined as the radius at which the spin angular velocity ($\Omega_s$ = 2$\pi$/P$_s$) of the pulsar is equal to the Keplerian angular velocity ($\Omega_k$ = $\sqrt{GM_*/r^3}$) of the material being accreted. Assuming a standard pulsar mass of 1.5M$_{\odot}$ and using the observed XTE J1829-098 rotation period (P$_s$) of 7.8 s, we get,

\begin{equation}
R_{co} = \left(\frac{GM_*}{4\pi^2} \times P_s^2\right)^\frac{1}{3} \simeq 6 \times 10^8 \rm cm.
\end{equation}

The Alfven radius (R$_{Alf}$) is defined as the radius where the ram pressure of the infalling material from the companion star ($\rho v^2$) balances with the magnetic pressure of the pulsar magnetosphere ($B^2/8\pi$). Assuming typical values for a pulsar, mass of 1.5$M_{\odot}$ and radius R$_*$ = 10$^6$ cm, the observed magnetic field of B $\simeq$ 1.7 $\times$ 10$^{12}$ G, resulting in a magnetic moment, $\mu$, of B R$_*^3$ $\simeq$ 1.7 $\times$ 10$^{30}$ G cm$^3$ and observed X-ray luminosity L$_{X} \simeq$ 4.3 $\times$ 10$^{36}$ erg/s \citep{shtykovsky_18}, we get the Alfven radius as \citep{lamb73, becker12},

\begin{equation}\label{eq2}
R_{Alf} = 2.6 \times 10^8 \left(\frac{\Lambda}{1}\right) \left(\frac{M_*}{M_\odot}\right)^\frac{1}{7} \left(\frac{R_*}{10^6\:\rm cm}\right)^{-\frac{2}{7}} \left(\frac{L_{X}}{10^{37}\:\rm erg/s}\right)^{-\frac{2}{7}} \left(\frac{\mu}{10^{30}\:\rm G\:\rm cm^3}\right)^\frac{4}{7} \rm cm \simeq 5 \times 10^8 cm.
\end{equation}

where the constant $\Lambda$ signifies the geometry of the accretion flow. Following \cite{becker12}, there is an uncertainty on the value of $\Lambda$, which is $\Lambda$ = 1 for spherical accretion, and $\Lambda$ $<$ 1 for disk accretion. Since very distinct accretion disks usually do not form in case of HMXBs \citep{karino19, Reig11}, in this letter we assume a wind-fed spherical accretion ($\Lambda$ = 1) for simplicity. R$_{Alf}$ for spherical accretion as given in equation \ref{eq2}, is very close to R$_{co}$, making this a case for intermittent accretion. In this regime, a turbulent and magnetized transition zone can be formed close to R$_{Alf}$, due to the balance between the magnetic pressure and the pressure inserted by accreting matter. Part of the infalling matter accumulated at the transition zone can further accrete onto the pulsar surface (accretor phase). However, the rotating pulsar magnetosphere can also strongly shock the infalling material at the trasition region, ejecting some of it beyond the accretion radius (propeller phase). Electrons can get shock-accelerated to very high energies at this transition region and can further produce X-rays via the synchrotron mechanism \citep{romanova15,torres_12, lovelace, bednarek}. Although X-rays produced from accretion are the dominant component observed during the outburst phase, a sub-dominant X-ray component at higher energies, produced from shocked electrons, should also be present in the data observed by NuSTAR during the same outburst phase. 

\begin{figure}[ht!]
\gridline{\fig{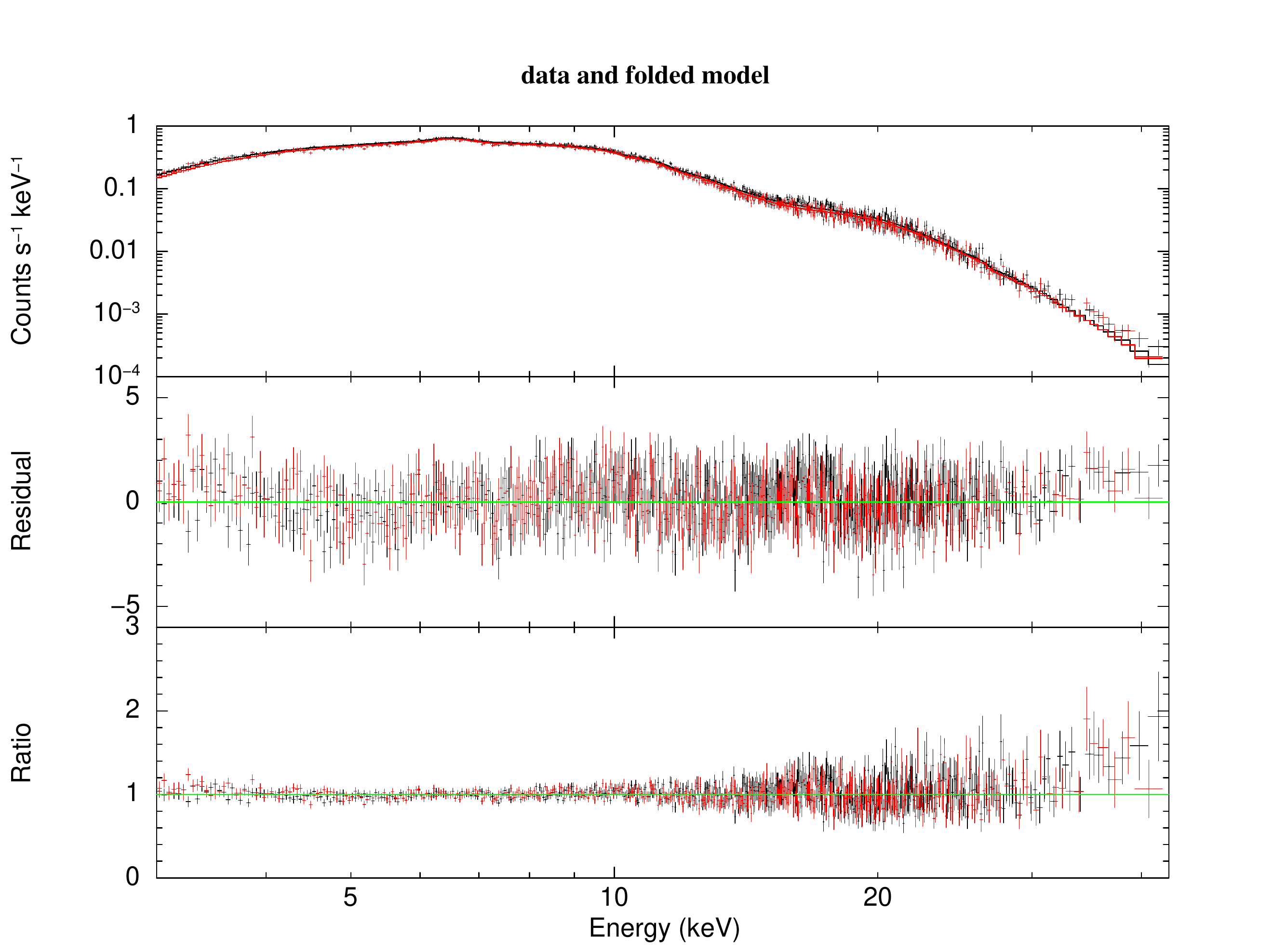}{0.5\textwidth}{(a)}
          \fig{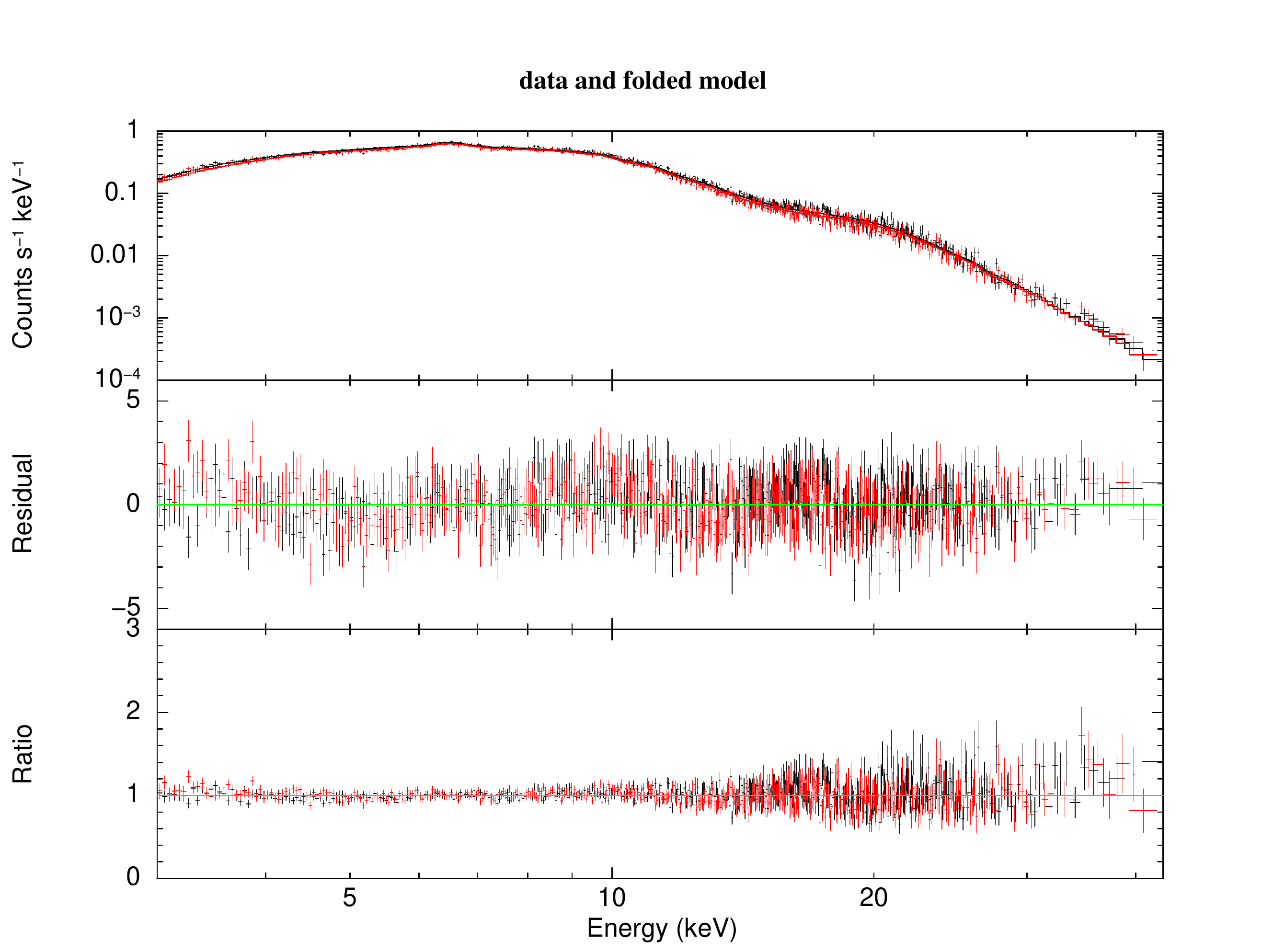}{0.5\textwidth}{(b)}
		 } 

	\caption{\label{fig1} (a) Data and model spectrum fit, the residual and the ratio (data/model) for the best-fit values given in the upper panel of Table \ref{tab1}. Model used in this case is \texttt{constant*tbabs*(cutoffpl*gabs + gauss)}. FPMA and FPMB data points and best-fits are shown in black and red respectively. (b) Data and model spectrum fit, the residual and the ratio (data/model), after addition of a power law component with the best-fit model used in (a). Model used in this case is \texttt{constant*tbabs*(cutoffpl*gabs + gauss + pow)}. Color scheme is the same as in (a).}
\end{figure}

To confirm this observationally, we have analyzed the data obtained by NuSTAR on 2018 August 16 (ObsID 90401332002), with an on-source exposure time of $\sim$ 27.8 ks and an average count rate of $\sim$ 8 cts s$^{-1}$ per module \citep{shtykovsky_18}. To extract the spectra, we have used the \texttt{NuSTAR-DAS} 2.0.0 software as distributed with the \texttt{HEASOFT} 6.28 package, with the CALDB version 20210315. The source data was extracted from a circular region of radius 50 arcsec, centered on the source position. The background data was extracted similarly from a circular region of radius 70 arcsec, away from the source position. The NuSTAR observations are not affected by stray light. The obtained spectra were grouped to have 25 counts per bin using \texttt{grppha} tool. The spectral analysis was done using the \texttt{XSPEC 12.11.1} tool included in the \texttt{HEASOFT} 6.28 package. Since the background starts to dominate the source counts above 50 keV, in this letter, we have considered 3 - 45 keV energy range for spectral analysis.

\begin{table*}[]
    \centering
    \begin{tabular}{cc}
    \hline
    \hline
        Parameter & Value \\
        \hline
        Hydrogen column density, N$_H$ (cm$^{-2}$) & 1.43 $\times$ 10$^{22}$\\
        Photon index of the cutoff power law, $\Gamma^{cutoffpl}_X$ & -0.75$^{+0.03}_{-0.03}$\\
        Folding energy of exponential rolloff, E$_{fold}$ (keV) & 4.49$^{+0.06}_{-0.06}$\\
        Cyclotron line energy, E$_{cyc}$ (keV) & 15.20$^{+0.10}_{-0.10}$\\
        Cyclotron line width, W$_{cyc}$ (keV) & 2.37$^{+0.10}_{-0.10}$\\
        Optical depth at Cyclotron line center, $\tau_{cyc}$ & 0.55$^{+0.05}_{-0.05}$\\
        Fe K$\alpha$ line energy, E$_{Fe}$ (keV) & 6.52$^{+0.04}_{-0.04}$\\
        Fe K$\alpha$ line width, $\sigma_{Fe}$ (keV) & 0.22$^{+0.04}_{-0.04}$\\
        \hline
        Photon index of the power law, $\Gamma^{pl}_X$ & 1.50$^{+0.15}_{-0.10}$\\
        \hline
    \end{tabular}
    \caption{\textit{Upper Panel} : Best-fit parameters of the model \texttt{constant*tbabs*(cutoffpl*gabs + gauss)}, along with their 1$\sigma$ uncertainties. \textit{Lower Panel} : Best-fit photon spectral index of the additional power law component, along with its 1$\sigma$ uncertainty.}
    \label{tab1}
\end{table*}

According to \cite{shtykovsky_18}, the spectrum of XTE J1829-098 can be explained by a power law with an exponential cutoff (\texttt{cutoffpl} model), modified by the fluorescent iron emission line (Gaussian line profile model \texttt{gauss}) and an absorption line (Gaussian absorption line model \texttt{gabs}), which is interpreted as a Cyclotron Resonant Scattering Feature (CRSF). So we have analyzed the phase-averaged NuSTAR data and tried to fit the spectrum with the model \texttt{constant*tbabs*(cutoffpl*gabs + gauss)}, representing the accretion component. We have used the \texttt{tbabs} model to take into account the X-ray absorption by the interstellar medium (ISM). To keep the best-fit values of the model consistent with the best-fit results obtained by \cite{shtykovsky_18}, we have kept the value of the hydrogen column density N$_H$ in the direction of XTE J1829-098, fixed at 1.43 $\times$ 10$^{22}$ cm$^{-2}$ \citep[LAB survey;][]{kalberla_05}. We have used atomic cross-sections from \cite{verner_96} and elemental abundances from \cite{wilms_2000}. The best-fit values along with their 1$\sigma$ uncertainties ($\chi^2$/D.O.F. = 1196.19/1071 $\approx$ 1.12), are shown in the upper panel of Table \ref{tab1}. Considering the uncertainties, the measured values of the model are consistent with those given in \cite{shtykovsky_18}. The flux obtained from the model in the 3 - 79 keV energy range was found to be F$^{acc}_X$ $\simeq$ (3.66 $\pm$ 0.02) $\times$ 10$^{-10}$ erg cm$^{-2}$ s$^{-1}$. The spectrum fit, along with residual and data/model ratio are shown in Figure \ref{fig1} (a). Although the best-fit values give a very good fit at low and intermediate energies, the best-fit model deviates from the data at higher energies, which is evident from the residual and ratio plots. This discrepancy hints towards a second emission component from the same source region. 

Next, we have added an additional power law spectrum, in the form of the model \texttt{pow}, with the above model signifying accretion, to fit the data. We have let the parameters of the power law component to freely vary, while keeping the best-fit values given in the upper panel of Table \ref{tab1} fixed. The best-fit photon spectral index value of the additional power law is given in the lower panel of Table \ref{tab1}. As found in other established HMGBs, the spectral index of the power law can vary between 1.4 and 1.6 \citep{takahashi_09}. It can be readily seen that the best-fit value along with the uncertainty of the additional power law component spectral index agrees well with previous observations. The obtained data and the corresponding best-fit model, along with the residual and the ratio, after fitting the data with the model \texttt{constant*tbabs*(cutoffpl*gabs + gauss + pow)}, are shown in the Figure \ref{fig1} (b). From the figure, it can be seen that the data is fitted comparatively well at higher energies after the addition of the power law model ($\chi^2$/D.O.F. = 1187.15/1076 $\approx$ 1.10). The absorbed flux of the sub-dominant power law component in the energy range of 3 - 79 keV was found to be F$^{pl}_X \simeq$ (9.6 $\pm$ 0.8) $\times$ 10$^{-12}$ erg cm$^{-2}$ s$^{-1}$, and the corresponding luminosity is L$_X^{pl}$ $\simeq$ (1.1 $\pm$ 0.1) $\times$ 10$^{35}$ (d/10 kpc)$^2$ erg s$^{-1}$. 

It was found that the improvement in the $\chi^2$ statistic after the addition of the sub-dominant power law component with the accretion component, is small. We have also calculated the F-statistic probability using \texttt{ftest} tool present in \texttt{XSPEC}. We have used appropriate $\chi^2$ and D.O.F. values for the calculation, and found that the F-statistic probability ($\approx 1 \times 10^{-2}$), although $<<$ 1, is comparatively high. These results suggest that the addition of the sub-dominant power law component with the accretion component is, although reasonable, of low statistical significance. This is not surprising as the additional power law component is sub-dominant compared to the dominant accretion component in the outburst phase of the XTE source. Moreover, the marginal improvement in the fit statistics can be attributed to a low number of datapoints available to constrain the additional power law component in the hard X-ray range. Nevertheless, the improvement in the residual and the ratio associated with the data and model X-ray spectrum (see Figure \ref{fig1} (a) and (b)), justify the addition of the sub-dominant power law component. Observational detection of this power law component, in conjunction with the argument presented above in terms of different characteristic radii, suggest that X-rays produced from shocked electrons through synchrotron cooling, are also present in the source region. In Appendix \ref{append:A}, we present the significance of the sub-dominant power law component, obtained using Monte Carlo simulation method. We note that calculating R$_{Alf}$ with $\Lambda$ = 0.5, as what may be expected from disk-fed accretion, yields an Alfven radius of R$_{Alf} \simeq$ 2 $\times$ 10$^8$ cm, which is, although of the same order, somewhat less than R$_{co}$. This may imply that the infalling material from the companion star accretes on the pulsar surface, without being propelled at the transition region. Consequently, no shock is created at the transition region in case of disk-fed accretion. However, signature of the shock component is observed in the NuSTAR data, represented by the sub-dominant power law component, indicating that our assumption of a wind-fed spherical accretion is valid. The presence of the sub-dominant, non-thermal power law emission indicates that this source indeed shows typical characteristics of a HMGB \citep{volkov_21, takahashi_09, an_15}. We have also performed pulse phase-resolved spectroscopy of the observed NuSTAR data in four different phase bins of equal sizes, spanning the entire phase range of 0 - 1, using the same model described above. But due to relatively low source photon counts, as well as large uncertainties associated with the datapoints, the phase dependence of the sub-dominant power law component could not be unambiguously established. Multiple simultaneous X-ray observations can help elucidate the phase dependence of the shock component.

\subsection{GeV counterpart of HESS J1828-099} \label{subsec:FLA}

Despite being very prominent in TeV energies, HESS J1828-099 has not been properly identified in GeV energies. For a deeper search of its GeV counterpart, we have analyzed $\sim$ 12 years of $\textit{Fermi}$-LAT data, observed between 2008 August 4 (MJD 54682) and 2020 October 2 (MJD 59124) in the 0.3–500 GeV band. A full description of the GeV data analysis is presented in Appendix \ref{append:B}. The closest GeV source is 4FGL J1830.2-1005, which was detected at a best-fit position of R.A. = 277.5300$^\circ$ $\pm$ 0.0342$^\circ$, and Decl. = -10.0730$^\circ$ $\pm$ 0.0262$^\circ$, only 0.292$^\circ$ away from the centroid of the H.E.S.S. source. 4FGL J1830.2-1005 was detected with a TS value of 458.53 and its spectral shape is logparabolic, expressed by the form, 
\begin{equation}
\rm\frac{dN}{dE} = N_0 \left(\frac{E}{E_b}\right)^{\rm-(\alpha_{GeV} + \beta_{GeV}log\left(\frac{E}{E_b}\right))}
\end{equation}

The best-fit parameters are $\alpha_{GeV}$ = 3.491 $\pm$ 0.011, $\beta_{GeV}$ = 0.7651 $\pm$ 0.0059, $E_b$ = 1.396 GeV. The average energy flux of this source is F$^{GeV}_{\gamma}$ = (1.88 $\pm$ 0.02) $\times$ 10$^{-5}$ MeV\,cm$^{-2}$\,s$^{-1}$. This flux is included in the spectral energy distribution shown in Figure \ref{fig3}.

We have analyzed the extension of the 4FGL J1830.2-1005 using RadialDisk and RadialGaussian models as templates. Fitting the extension with the RadialDisk template gives a maximum TS$_{ext}$ value of 32.41 ($\sim$ 5.692$\sigma$), with best-fit 68$\%$ containment radius of the disk being 0.325$^\circ$ $\pm$ 0.037$^\circ$. We have considered radial disks of radius varying from 0$^\circ$ to 0.5$^\circ$ to show how the delta log-likelihood varies with increasing radius (see Figure \ref{fig2} (b)).  We have also studied the energy dependent morphology of the source by estimating the extent in two different energy ranges, 0.3 - 1 GeV and 1 - 500 GeV. We found that the spatial extent in both cases remains almost the same, 0.3063$^{+0.0630}_{-0.0692}$ degree in 0.3 - 1 GeV range and 0.2875$^{+0.0517}_{-0.0463}$ degree in 1 - 500 GeV range. It was found that the offset in spatial position of the 4FGL source at different energy ranges varies significantly from the original 4FGL source position (offset $\approx$ 0.1068$^\circ$ in the energy range 1 - 500 GeV and offset $\approx$ 0.0198$^\circ$ in the energy range 0.3 - 1 GeV).  The energy dependent morphology of the sources is shown in Figure \ref{fig2} (a). From the figure, it can be seen that the 4FGL source and the H.E.S.S. source overlap with each other. Also with increasing energy (in 1 - 500 GeV range), we observe an increment in spatial proximity between the 4FGL and the H.E.S.S. sources. Based on the positional coincidence between these two sources, it can be inferred that 4FGL J1830.2-1005 can possibly be the GeV counterpart of HESS J1828-099. A periodicity search was carried out to probe any possible periodic variation in the GeV $\gamma$-rays from the 4FGL source, but no significant periodicity was found. Details of the periodicity search is presented in Appendix \ref{append:C}.

\subsection{Radio counterpart of HESS J1828-099} \label{subsec:RDA}

We have used multiwavelength radio data from different surveys to look for possible counterparts of HESS~J1828-099. The field is observed as a part of the recent high sensitivity Galactic plane surveys like THOR survey \citep[the HI/OH/Recombination line survey;][]{thor1,thor2} covering $1-2$ GHz and the GLOSTAR Galactic Plane survey \citep[A GLObal view of STAR formation][]{glostar1} covering $4-8$ GHz. Due to the proximity of the source to the Galactic plane, the field is crowded with multiple resolved and unresolved sources including Galactic (H~{\sc ii} regions, supernova remanants, planetary nebulae) as well as many unclassified Galactic as well as extragalactic sources \citep{arnab20}. Near the position of XTE J1829-098, we detect a radio source within 99\% RXTE confidence region in both THOR and the GLOSTAR images, and investigate this as a plausible radio counterpart of the HMXB based on its proximity. A full description of the radio data analysis is presented in Appendix \ref{append:D}.

Figure~\ref{fig2} (c) shows the 1.4 GHz THOR+VGPS image of the field at $25\arcsec$ resolution. There is no radio emission at the Chandra position of XTE~J1829-098. However, within the RXTE error region, marked by the ellipse with cross-hair at the centre, there is a prominent radio source detected in THOR. The source is marginally resolved and the L-band peak flux density of this source at an effective frequency of 1.63 GHz is $4.15 \pm 0.25$ mJy/beam. 

\begin{figure}[ht!]
\gridline{\fig{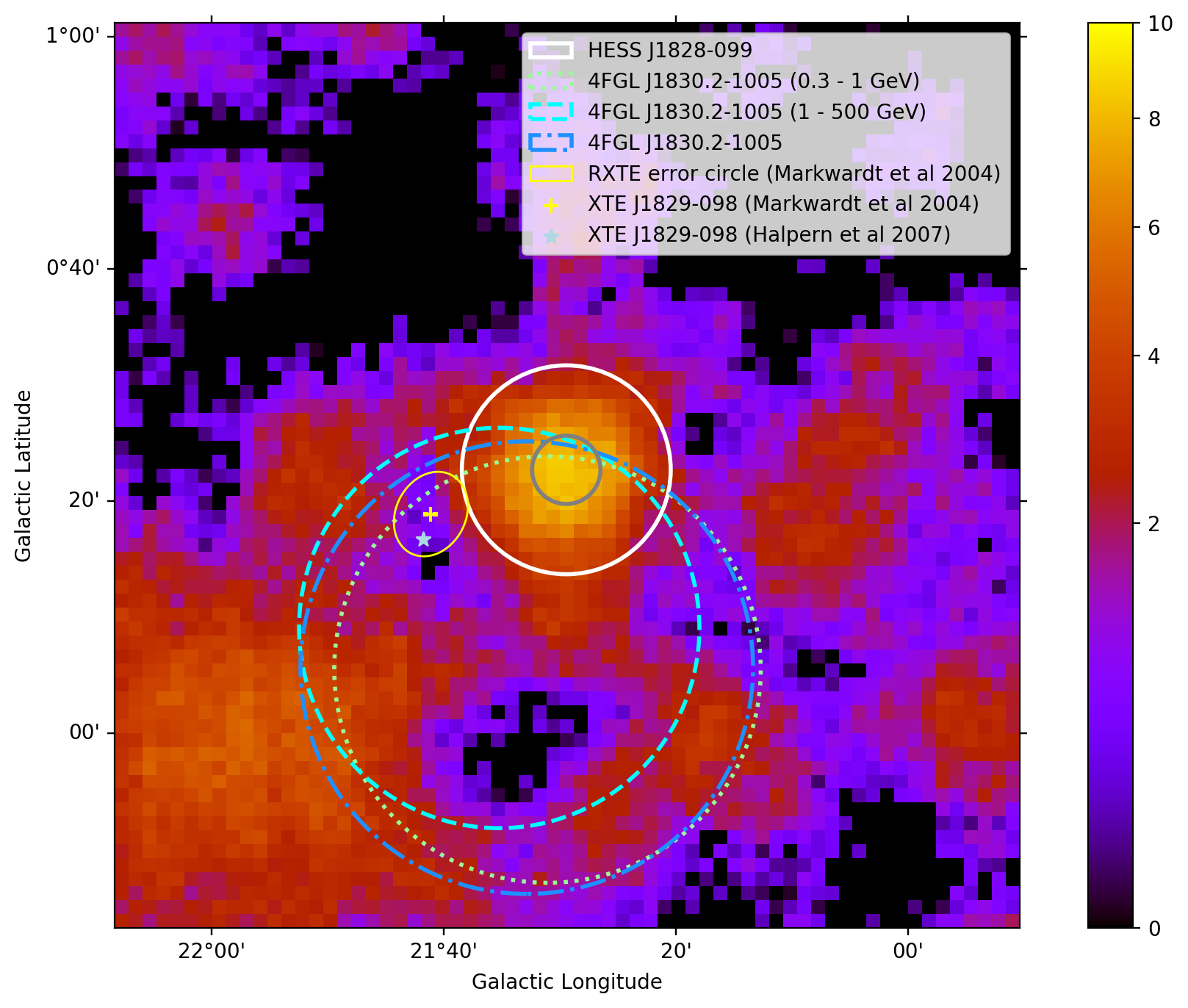}{0.5\textwidth}{(a)}
          \fig{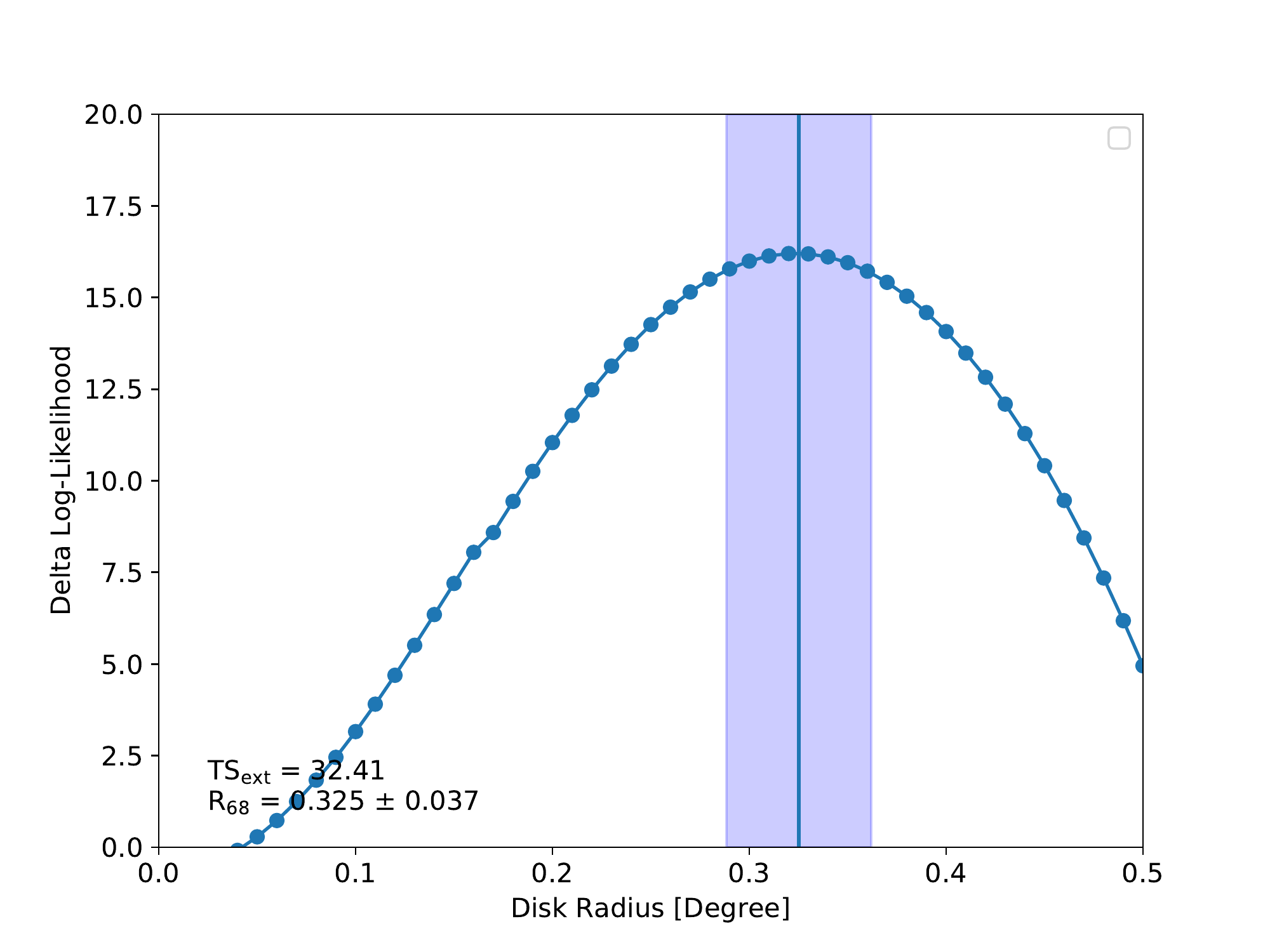}{0.5\textwidth}{(b)}
         }  
\gridline{\fig{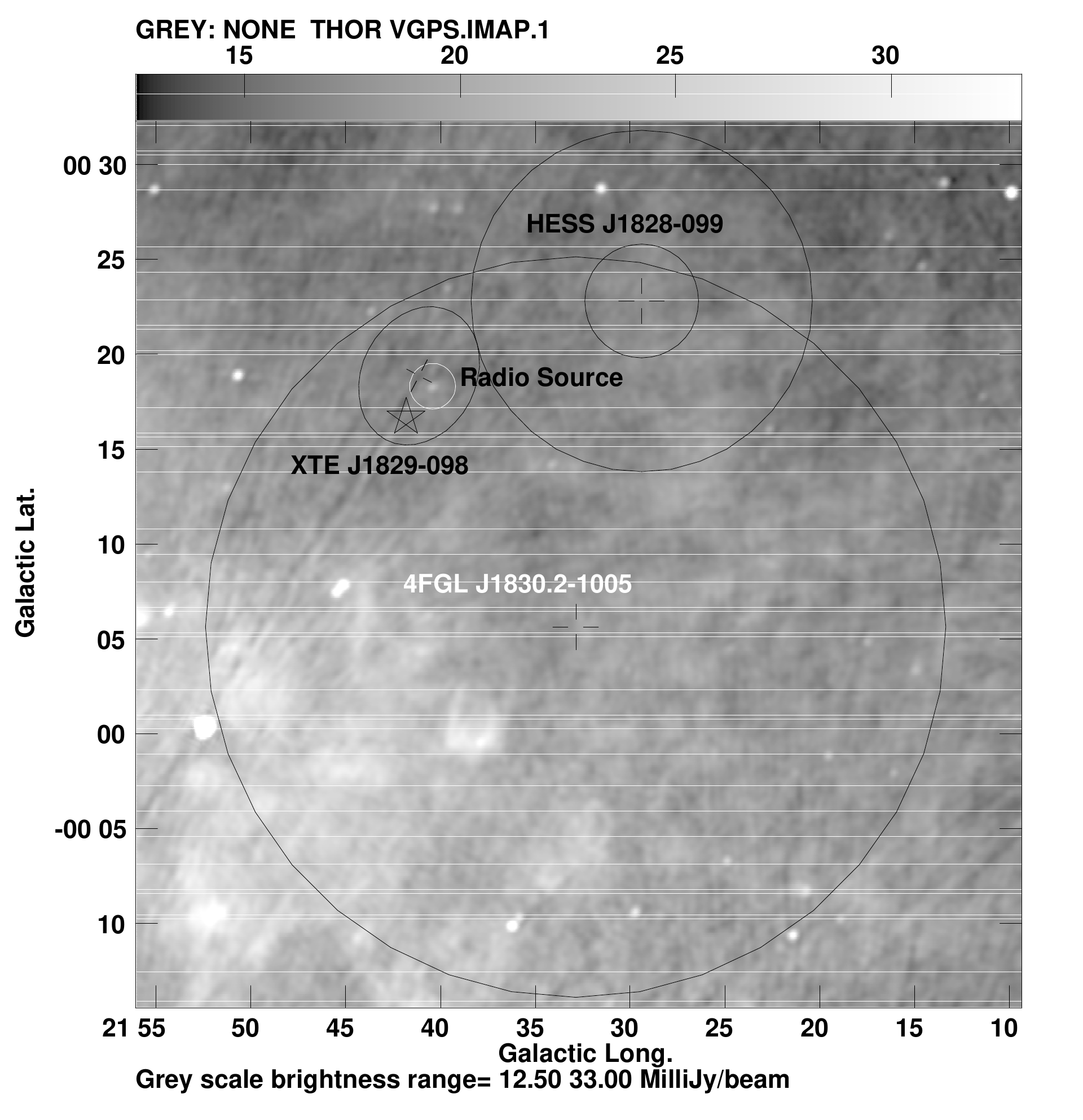}{0.5\textwidth}{(c)}
          \fig{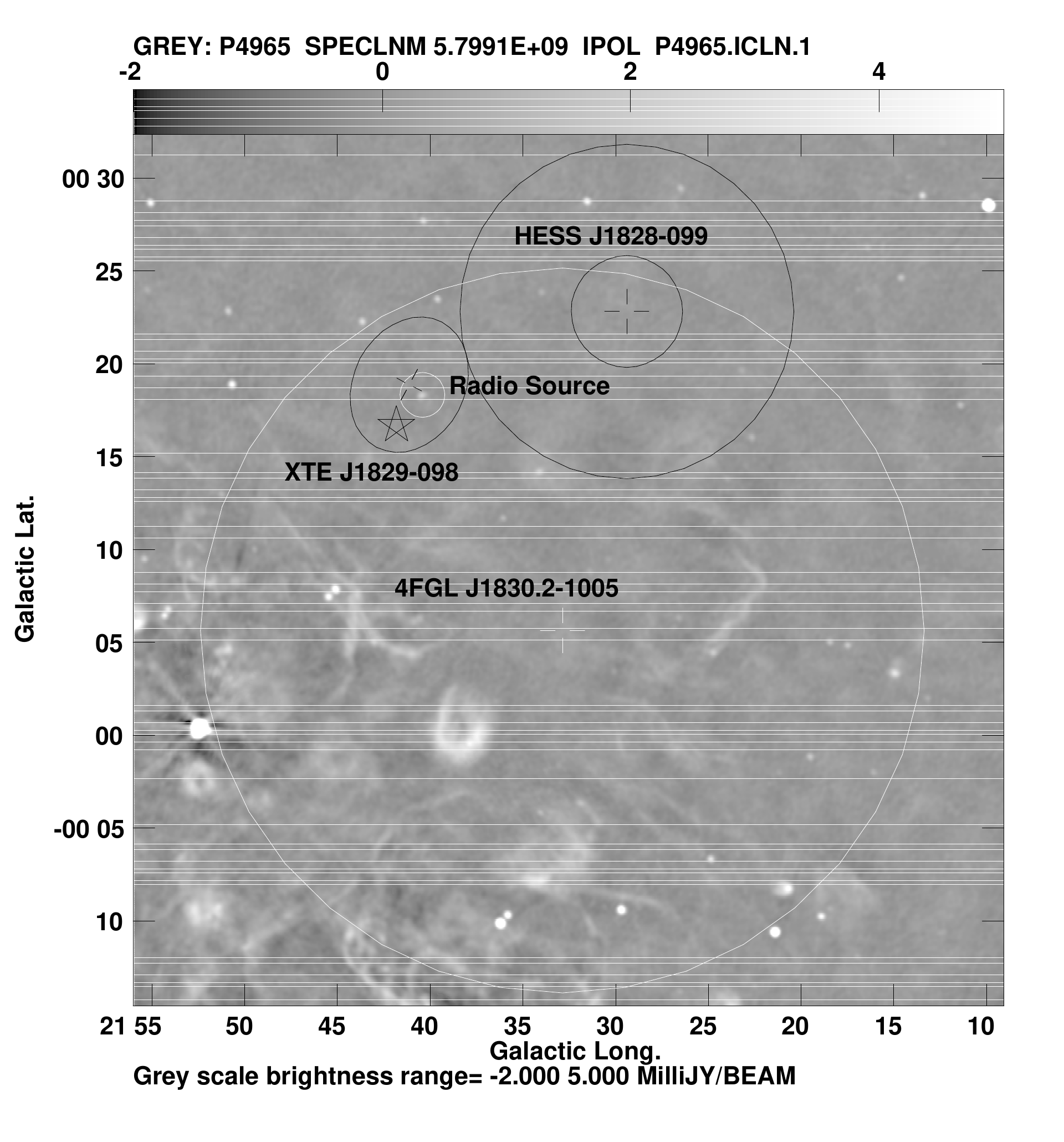}{0.5\textwidth}{(d)}
         }  
	\caption{\label{fig2} (a) H.E.S.S. significance map centered on HESS J1828-099. The colorbar denotes $\sqrt{TS}$ value of the region. The grey circle represents the extent up to which a 1D gaussian template was fitted and the white circle signifies the region within which spectral points for HESS J1828-099 were extracted. Morphologies of 4FGL J1830.2-1005 at different energy ranges are shown with green dotted line (0.3 - 1 GeV) and cyan dashed line (1 - 500 GeV). Blue dot-dashed line signifies spatial extension of the 4FGL in the entire considered energy range (0.3 - 500 GeV). RXTE position of pulsar XTE J1829-098 \citep{halpern_07}, alongwith 99$\%$ confidence region \citep{markwadt_04} are also shown in yellow. The Chandra position of the pulsar is shown with a light-blue star,(b) Variation of the delta log-likelihood value of 4FGL J1830.2-1005 modelled with radial disks of different radii. The blue shaded region indicates the uncertainty estimate of the best-fit extension of 4FGL J1830.2-1005. (c) The combined THOR and VGPS 1.4 GHz image and (d) the GLOSTAR 5.8 GHz image showing the radio continuum emission from the field containing HESS~J1828-099, 4FGL~J1830.2-1005, and the pulsar XTE J1829-098. The Chandra position of the pulsar is marked with a star and the RXTE error region is shown with a black ellipse. Spatial extent marked for the H.E.S.S. and the 4FGL sources (0.3 - 500 GeV) are same as in (a). Plausible radio counterpart of the binary system is marked by a white circle.}
\end{figure}

 The source identified from THOR as the possible counterpart of the binary system, marked by small white circle in Figure \ref{fig2} (c), is also detected in the GLOSTAR survey and has a peak flux density of $2.30 \pm 0.21$ mJy/beam (Figure \ref{fig2} (d)). The flux density values from the GLOSTAR subimages are consistent with the in-band spectral index ($\alpha_{\rm radio}$ where $S_\nu \propto \nu^{\alpha_{\rm radio}}$) of $- 0.746 \pm 0.284$ estimated from the flux values in different THOR spectral windows (SPWs). The observed radio spectrum is indicative of particle acceleration due to the collision of an ultrarelativistic pulsar wind and the wind/disk of the normal star. The extended nature of the source indicates its possible Galactic origin. In the complete catalog of the D configuration continuum sources (Medina et al., in prep.), it is classified as a candidate planetary nebula based on its Mid-IR properties. However, the non-detection of this source in the earlier 1.4 GHz NVSS image \citep[the NRAO VLA Sky Survey;][]{condon_98} at $45\arcsec$ resolution also indicate variability of this source. We note that the putative radio source is also detected at 147.5 MHz in the TIFR GMRT Sky Survey \citep[TGSS;][]{tgss} with a flux density of $51.14 \pm 8.45$ mJy. The SED in Figure~\ref{fig3} includes the multiwavelength radio data from the TGSS, THOR SPWs as well as from the GLOSTAR subimages. Considering the spectral index, possible variability and the position of the source (within the RXTE error region but not coinciding with the Chandra position of XTE~J1829-098), in the subsequent analysis we consider both the possibilities that this radio source may or may not be a counterpart of HESS~J1828-099. For the scenario that it is not associated, we have used the $3\sigma$ limits from the GLOSTAR, THOR and the TGSS to construct (and model) the SED.

\section{Multi-wavelength SED modelling} \label{SED_modelling}

\begin{figure}[ht!]
\gridline{\fig{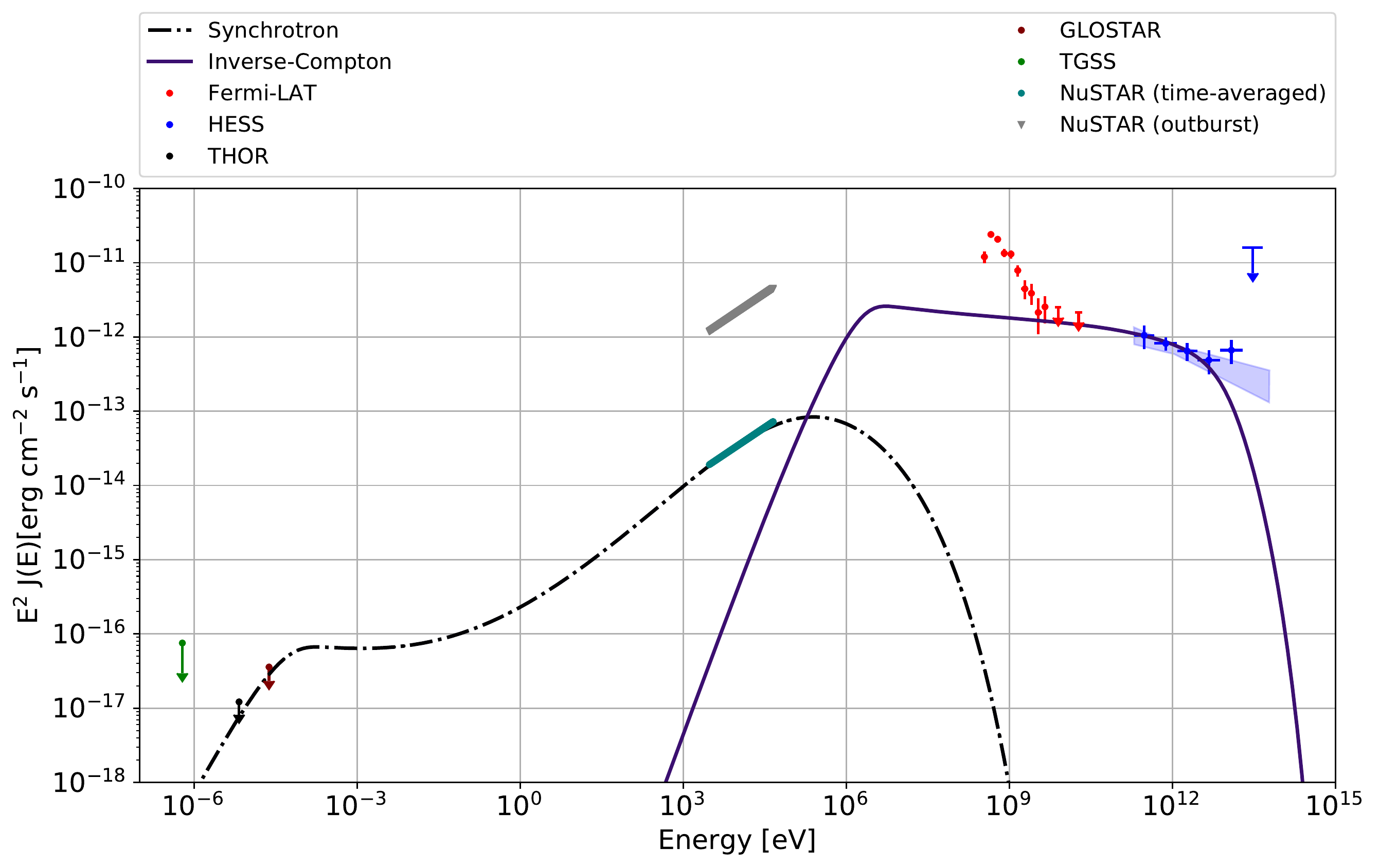}{0.5\textwidth}{(a)}
          \fig{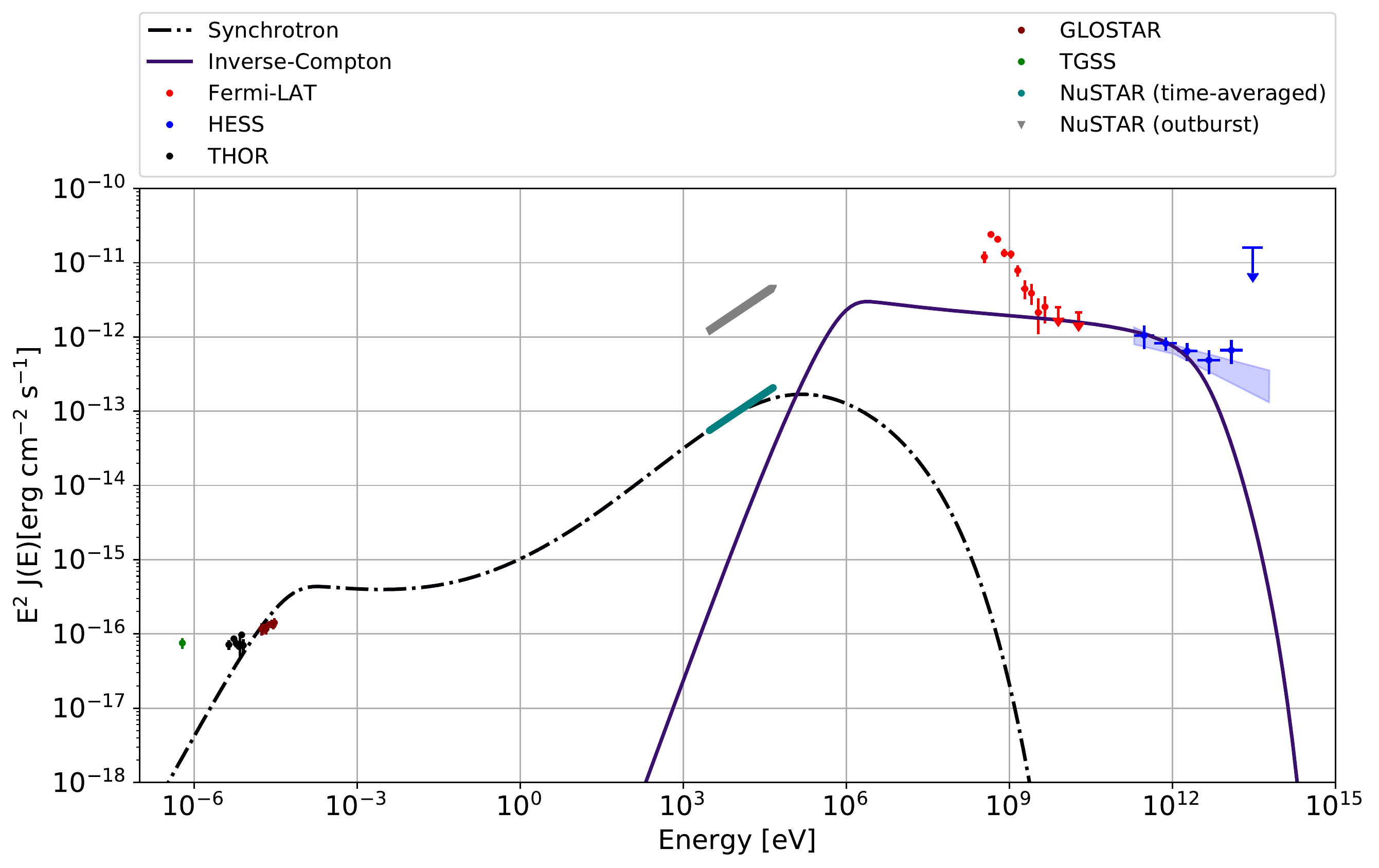}{0.5\textwidth}{(b)}
          }
\gridline{\fig{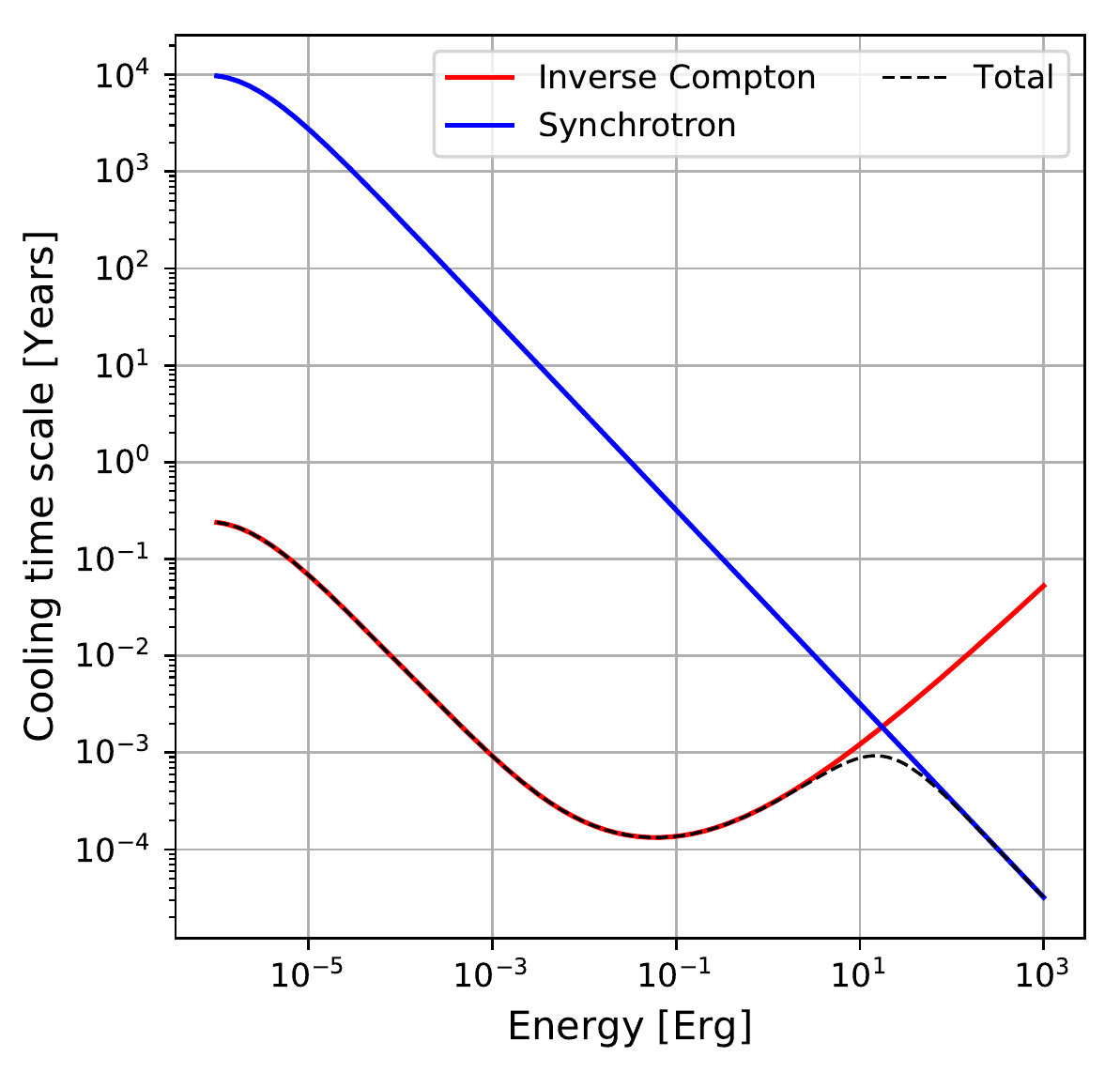}{0.3\textwidth}{(c)}
          \fig{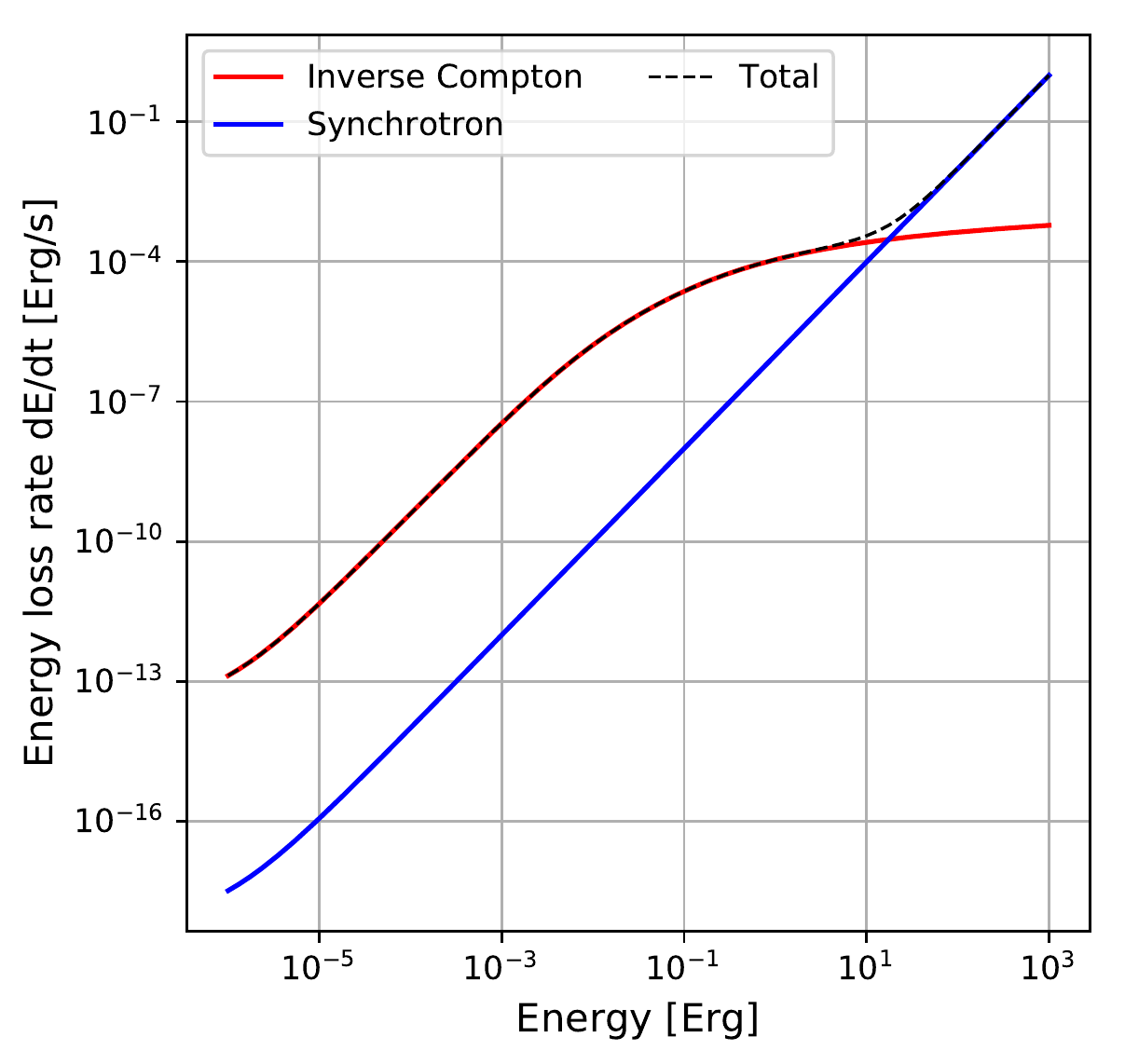}{0.3\textwidth}{(d)}
          }
\gridline{\fig{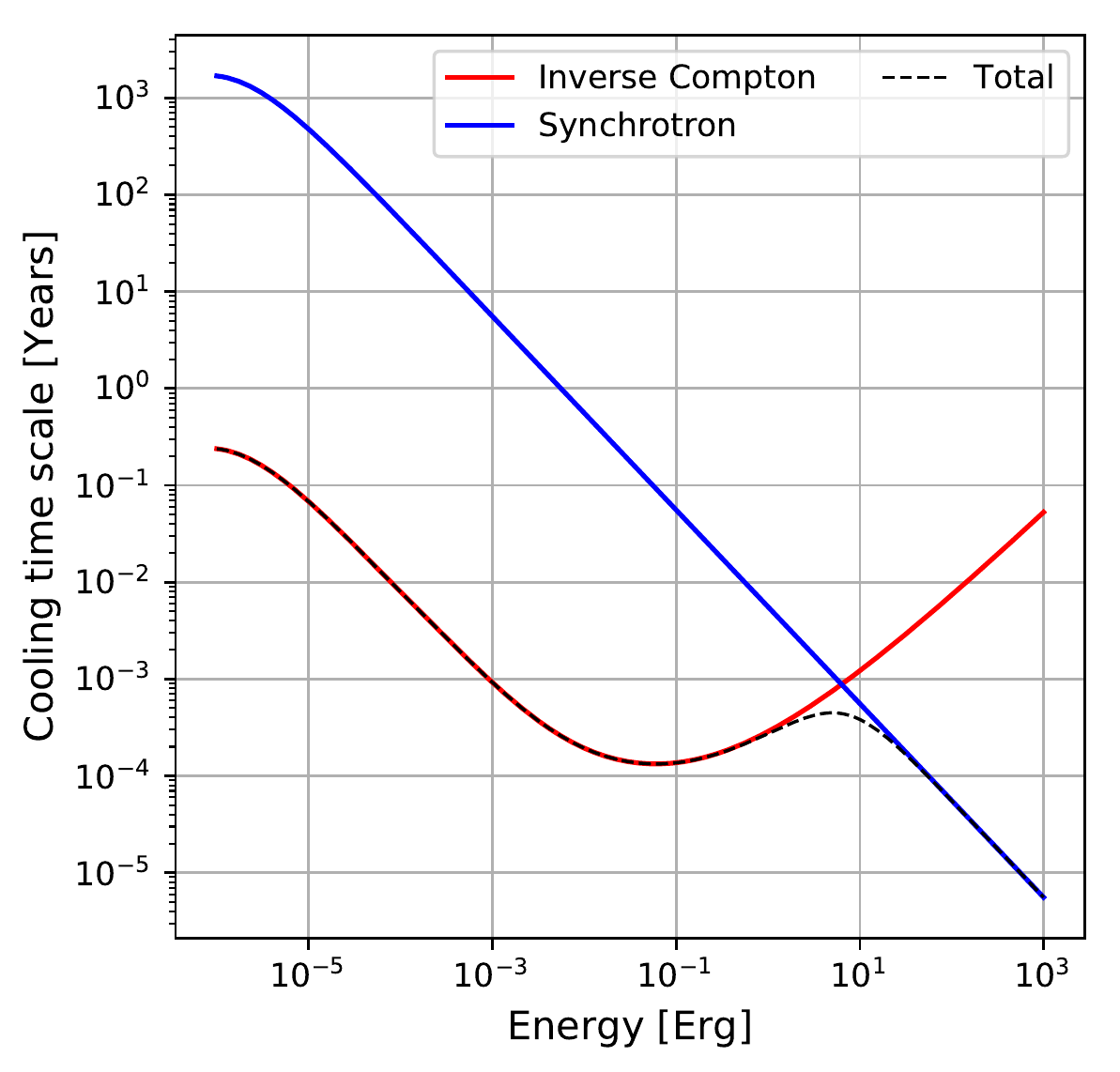}{0.3\textwidth}{(e)}
          \fig{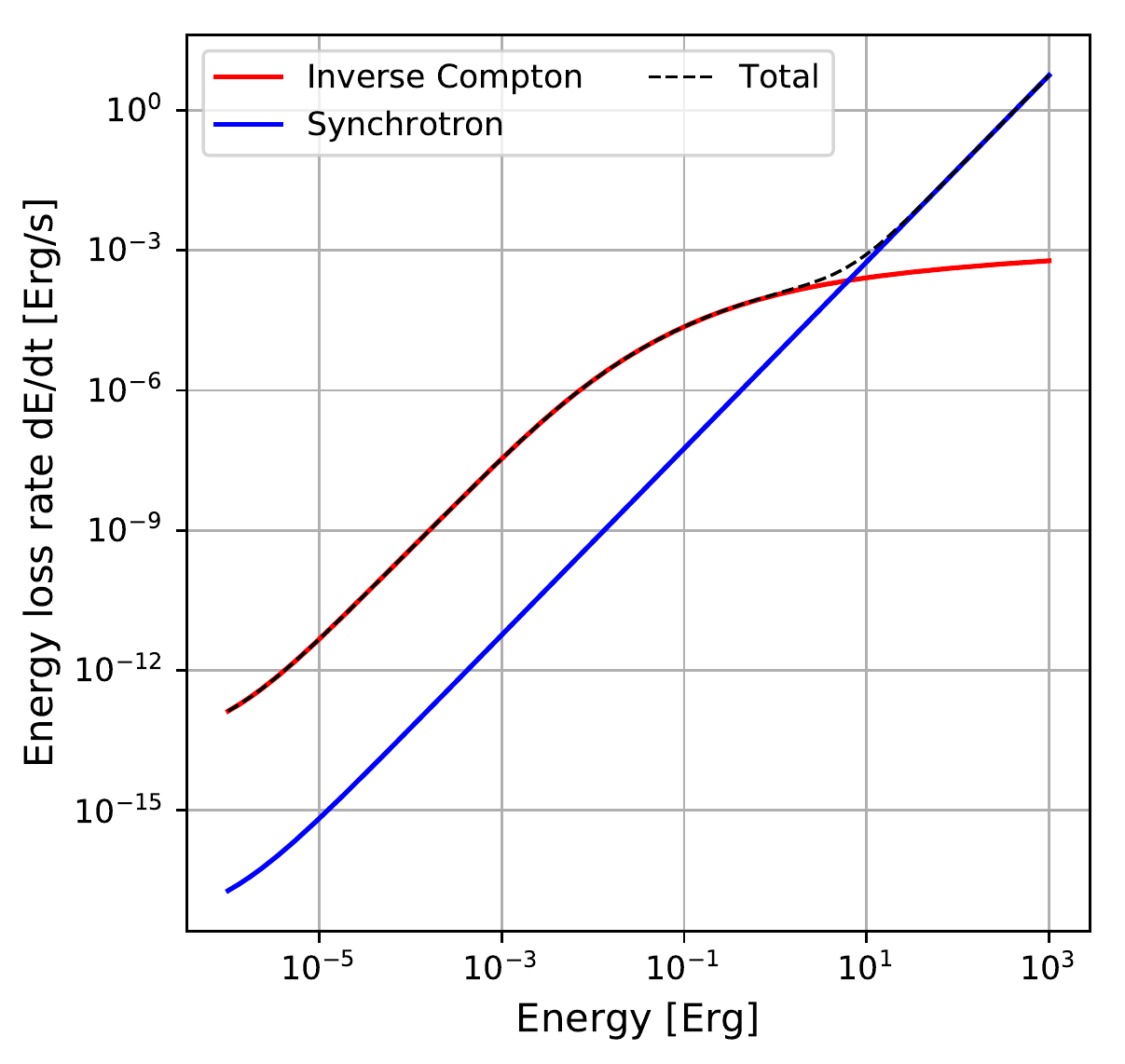}{0.3\textwidth}{(f)}
          }          
\caption{\label{fig3} Multiwavelength SED of the source HESS J1828-099 and corresponding IC dominated (a) model 1 and (b) model 2, obtained using GAMERA. The unabsorbed power law X-ray SED obtained from NuSTAR data analysis in the outburst phase of XTE J1829-098, is shown with grey datapoints. The same unabsorbed X-ray SED, time-averaged over the orbital period of XTE J1829-098 \citep{halpern_07}, is shown with teal datapoints. The H.E.S.S. data, shown in blue, was taken from \cite{HESS_18_2}. We have analysed the \textit{Fermi}-LAT data and the corresponding SED from 4FGL J1830.2-1005 is shown in red. 3$\sigma$ upper limits at radio range, obtained at the Chandra position of XTE J1829-098, observed by THOR (black), GLOSTAR (maroon) and TGSS (green), are shown in (a) with downward arrows. In (b), flux values of the putative radio source from these surveys are shown with the same colour scheme. In (c) and (d), we present the cooling timescale and energy loss rate of model 1, at time t = t$_{age}$ $\approx$ 10$^7$ years. In (e) and (f), we plot the same as (c) and (d), for model 2.}
\end{figure}

We have accumulated the data obtained from different multi-wavelength observations, shown in Figure \ref{fig3} (a) and (b), to perform multi-wavelength SED modelling. We have considered a leptonic, Inverse Compton (IC) dominated, one-zone model, similar to \cite{hinton_08}, to explain the emission from HESS J1828-099. Since there is an offset between the Chandra position of XTE J1829-098 and the putative radio source found in the RXTE error region, we have explored two different cases to explain the multi-wavelength SEDs. In Model 1, we consider 3$\sigma$ upper limits for radio flux density at the exact Chandra position of XTE J1829-098 and use these upper limits to construct the SED at radio frequencies, whereas in Model 2, the radio source within RXTE error region is assumed to be the radio counterpart of the HMXB and the GLOSTAR/THOR/TGSS data are used to extend the SED to radio wavelengths.

The HMXB XTE J1829-098 is located at a distance of 10 kpc from Earth \citep{halpern_07}. Since the companion star of the HMXB probably is a Be star, we assume its age is t$_{age}$ $\le$ 10$^7$ years and the stellar photon temperature T$_{*}$ is $\approx$ 30000 K \citep{Takata_2017}. 
We have considered a population of accelerated electrons having a cutoff power law spectrum, dN/dE$_e$ $\propto$ E$_e^{-\alpha_e}$\,exp(-E$_e$/E$_{max}$) in the shock region between the pulsar and the companion star. Small distance between the companion star and the pulsar ($\sim$ 0.2$''$) indicates that a photon field with high radiation density is present in the region.  The ultra-relativistic electrons are cooling down by synchrotron and IC emission. Radio to X-ray emission is produced due to synchrotron emission and $\gamma$ rays are produced by IC emission. As discussed in subsection \ref{subsec:XRAY}, we detected a sub-dominant, power law X-ray component with a spectral index of 1.50$^{+0.15}_{-0.10}$, which implies that the energy spectrum of parent electrons should have a power law spectral index $\alpha_e$ = 2$\Gamma^{pl}_{X}$ -1 = 2.0$^{+0.3}_{-0.2}$. We have searched within this range to find the best-fit spectral index for the parent electron spectrum for both model 1 and model 2. Moreover, we have also used an exponential cutoff in the parent electron spectrum, as electron being leptons, lose energy very efficiently. We have assumed E$_{max}$ = 50 TeV, maximum energy upto which the parent electrons can be accelerated in the shock site.

\begin{deluxetable}{ccccccccc}
\tablecaption{\label{tab2} Parameters used for two models}
\tablewidth{0pt}
\tablehead{
\colhead{Model} & \colhead{E$_{min}$} & \colhead{E$_{max}$} & \colhead{$\alpha_e$} &
\colhead{B} & \colhead{$T_{*}$} & \colhead{U$_{rad}$} & \colhead{Age} & \colhead{Distance} \\
\nocolhead{-} & \colhead{(GeV)} & \colhead{(GeV)} & \nocolhead{-} & \colhead{(mG)} &
\colhead{(K)} & \colhead{(erg\,cm$^{-3}$)} & \colhead{(Years)} & \colhead{(kpc)}
}
\startdata
Model 1 & 0.12 & 5$\times$10$^4$ & 2.2 & 25 & 30000 & 1 & 10$^7$ & 10\\
\hline
Model 2 & 0.08 & 5$\times$10$^4$ & 2.2 & 60 & 30000 & 1 & 10$^7$ & 10\\
\enddata
\end{deluxetable}

By analyzing the NuSTAR data, the fluxes of the accretion component (F$_X^{acc}$ $\simeq$ (3.66 $\pm$ 0.02) $\times$ 10$^{-10}$ erg cm$^{-2}$ s$^{-1}$) and the shock component (F$_X^{pl}$ $\simeq$ (9.6 $\pm$ 0.8) $\times$ 10$^{-12}$ erg cm$^{-2}$ s$^{-1}$) in 3 - 79 keV range, during the outburst phase, were determined, as discussed in subsection \ref{subsec:XRAY}. But XTE J1829-098, being a transient source, shows a very high observed dynamic range ($\sim$ 6800) \citep{halpern_07}, which indicates that the value of F$_X^{acc}$ can decrease down to $\sim$ 10$^{-14}$ erg cm$^{-2}$ s$^{-1}$ in its most quiescent phase. The flux of the shock component F$_X^{pl}$ will also decrease when the XTE source is not in the outburst phase. Due to the lack of long-term observational data, we assume that the time-averaged flux of the shock component over the entire orbital periodic revolution, is (1 - 5) $\times$ 10$^{-2}$ times the flux measured in the outburst phase. 
This assumption is not unreasonable since the XTE source spends comparatively less time in the outburst phase during its orbital motion, making the time-averaged flux lower than that in the outburst phase. Moreover, other datasets in the radio, GeV and TeV ranges considered in this letter for multiwavelength SED construction, are collected from long-term observations, whereas the NuSTAR data for the XTE source is only observed during the outburst phase. Hence, to keep the multiwavelength SED modelling consistent, we have assumed time-averaged X-ray fluxes from the XTE source. The assumed time-averaged X-ray fluxes used for model 1 and model 2 in 3 - 79 keV range  are, F$^{pl, 1}_X$ $\simeq$ (1.5 $\pm$ 0.1) $\times$ 10$^{-13}$ erg cm$^{-2}$ s$^{-1}$ and F$^{pl, 2}_X$ $\simeq$ (4.4 $\pm$ 0.3) $\times$ 10$^{-13}$ erg cm$^{-2}$ s$^{-1}$ respectively. Although some uncertainties might be associated with the assumed X-ray flux values, the data is within the dynamic range of the XTE source, which future observations can verify. 

Previously, \cite{hinton_08}, modelled the multi-wavelength data of the TeV HMGB HESS J0632+057 using a one-zone leptonic model. We adopt the same value of the suppression factor due to KN effect from \cite{hinton_08}, i.e. f$_{KN}$(E$_e$) $\sim$ 10$^{-3}$ for kT$_*$ $\sim$ 3 eV and E$_e$ = 1 TeV. For this value of f$_{KN}$, the magnetic field was calculated from the relation B $\approx$ 5(f$_{KN}$F$_X$/F$^{TeV}_{\gamma}$)$^{0.5}$ G, where F$_X$ and F$^{TeV}_{\gamma}$ are fluxes of X-rays and TeV $\gamma$-rays \citep{HESS_18_2} respectively. We have considered a photon radiation density similar to that of \cite{hinton_08}, i.e. U$_{rad} \sim$ 1 erg\,cm$^{-3}$. The IC emission of ultra-relativistic electrons is happening in the deep Klein-Nishina (KN) regime \citep{hinton_08}, as a result the TeV $\gamma$ ray spectrum is softer compared to the X-ray spectrum produced by synchrotron emission.  Such spectral variation was seen in X-Ray and TeV ranges for our source \citep{halpern_07, HESS_18_2}, which is a characteristic feature of HMGBs.

We have  studied the radiation from synchrotron and IC cooling of ultra-relativistic electrons, by solving the particle transport equation using publicly available code GAMERA\footnote{\url{https://github.com/libgamera/GAMERA}} \citep{hahn_15}. We vary  the total injected power in electrons, to fit the multi-wavelength data of HESS J1828-099. The parameters required to explain the multi-wavelength data in both cases, are given in Table \ref{tab2}. Both model 1 and model 2, depicted in Figure \ref{fig3} (a) and (b) respectively, require a power of $\sim$ (4 - 5) $\times$ 10$^{35}$ erg\,s$^{-1}$. Although the multi-wavelength one-zone models fail to reproduce the spectrum in the GeV range in both cases, the required luminosity in electrons of the models and the required parameters shown in Table \ref{tab2}, are consistent with those of the firmly established TeV HMGBs, thus indicating that  HESS J1828-099 is possibly a TeV HMGB \citep{Takata_2017}. We also present the cooling time scale and energy loss rate of IC and synchrotron mechanisms considered in our models, in Figure \ref{fig3} (c) and (d) respectively for model 1, and in Figure \ref{fig3} (e) and (f) respectively for model 2.

\section{Discussion and conclusion} \label{sec:DAC}

The multiwavelength SED of HESS J1828-099 shown in Figure \ref{fig3} closely resembles that of other known TeV HMGBs, as all of the firmly established HMGBs have hard X-ray spectra and significantly softer spectra in TeV energies. Through a detailed \textit{Fermi}-LAT data analysis, the SED in the GeV energy range was also obtained. This type of spectral shape was seen previously in \cite{Tam_20}, who assumed that GeV emission is due to some unrelated source such as SNR G22.7-0.2, which is co-spatial with  HESS J1832-093 and 4FGL J1832.9-0913.

Since the resultant radiation from the hadronic \textit{p-p} interaction between protons accelerated in the SNR shocks and cold protons clumped in nearby clouds can explain the analysed GeV data, we have searched SNRs in the vicinity of HESS J1828-099. SNR G021.5-00.1, which has been detected in radio observations, was thought to be spatially coincident with 4FGL J1830.2-1005 \citep{hewitt_09, kilpatrick_15, acero_16}. Similarly, SNR G20.4+0.1, which is 1$^\circ$ away from HESS J1828-099, was assumed to be associated with the H.E.S.S. source \citep{pop_18}. However, it was found from THOR + VGPS data, as well as in GLIMPSE and WISE data, that these are clumped HII regions and not SNRs \citep{anderson_17}. Recently, in the GLOSTAR Galactic plane survey data, 4 SNR candidates were identified : G021.492-0.010, G021.596-0.179, G021.684+0.129 and G021.861+0.169, which fall within the positional uncertainty of 4FGL J1830.2-1005 \citep{dokara_2021}, however further observations are needed to establish a molecular cloud association with these SNRs. Alternatively, since 4FGL J1830.2-1005 is in a crowded region of Galactic plane, contamination from nearby pulsars can be significant. We tried to find any bright GeV $\gamma$-ray emitting pulsar in the 4FGL catalog, in the nearby region of 4FGL J1830.2-1005, but did not find any. If future observations detect a pulsar in the vicinity of the 4FGL source that is contaminating the GeV emission, then it might be possible to explain the GeV data by gating off the pulsar contribution using up-to-date ephemeris. At present, studying these scenarios is beyond the scope of this work. Our model 2 also fails to explain the TGSS data at 147.5 MHz (see Figure \ref{fig3}(b)). Since the HMXBs show strong variability in the X-ray range and the TGSS radio measurements were performed at a different epoch than the X-ray observations, radio variability can be a possible reason behind this discrepancy. Alternatively, a completely different non-thermal low energy radio component can also explain the TGSS data.  Simultaneous observations in the X-ray and radio ranges can help to address this discrepancy. While usually pulsars are the compact objects in HMGBs such as PSR B1259-63 and PSR J2032+4127, there is a recent debate on the nature of the compact object in LS 5039, which may actually be a magnetar with a spin period of 9 s \citep{volkov_21, yoneda_20, yoneda_21}. Although the spin period of the proposed magnetar is very close to the spin period of XTE J1829-098, the surface magnetic field of the magnetars is typically around 10$^{13}$ - 10$^{15}$ G, whereas for the compact object in this binary source, the magnetic field is lower compared to that ($\approx$ 10$^{12}$ G), confirming that the compact object in this HMXB system, is indeed a pulsar and not a magnetar.

Based on the definition of HMGBs \citep{Dubus_06, Dubus_13, Dubus_15, Dubus_17}, the emission typically dominates above 1 MeV. In the case of HESS J1828-099, the average GeV flux observed by \textit{Fermi}-LAT, F$_{\gamma}^{GeV}$($\simeq$ (3.01 $\pm$ 0.03) $\times$ 10$^{-11}$ erg\,cm$^{-2}$\,s$^{-1}$), is higher than the time-averaged X-ray flux values used both for model 1 and model 2, F$^{pl, 1}_X$ and F$^{pl, 2}_X$ respectively. Also from Figure \ref{fig3} (a) and (b), it can be seen that the multiwavelength SED peaks above 1 MeV. This nature of emission indicates that HESS J1828-099 can be classified as a HMGB. Furthermore, the required values of parameters presented in Table \ref{tab2}, resemble those of known TeV HMGBs \citep{skilton_09, hinton_08}. We have kept the distance of the HMXB source ($\sim$ 10 kpc) fixed \citep{halpern_07}. The environmental parameters such as magnetic field (B) and radiation density (U$_{rad}$) were assumed according to \cite{hinton_08}, and they were also kept fixed. Age (t$_{age}$) and stellar photon temperature (T$_*$) were consistent with the Be companion star \citep{Takata_2017}. The best-fit electron spectral index ($\alpha_e$) was calculated considering the uncertainty in the power law spectral index of the newly detected, sub-dominant, additional X-ray component, produced in the shock region between rotating pulsar magnetosphere and infalling stellar material. The magnetic fields used both for model 1 and model 2, are of the same order as in other established HMGBs \citep{hinton_08}, indicating that our assumption of the time-averaged X-ray flux is reasonable. The electron injection luminosity is the only free parameter which was varied to fit the data. Minimum energy of the parent electron population E$_{min}$ in model 1 is an upper limit, as the radio upper limits do not represent a detection themselves. Considering the offset between the Chandra position of XTE J1829-098 and the putative radio source, model 1 seems to be the favourable interpretation of the source, although model 2 is also plausible. Taking into account the fact that this HMGB is at a larger distance compared to other known binaries, the required electron injection luminosity is consistent with that reported for other established HMGBs \citep{Takata_2017, skilton_09, Eger_16}. 

In this paper, we have performed GeV, X-ray and radio data analyses and used results from previous infrared data analyses. From the X-ray data analysis, we have detected a sub-dominant, hard X-ray tail in the NuSTAR source spectrum of XTE J1829-098, which suggests that the X-rays are produced via synchrotron cooling of shocked electrons. However, alternate interpretations for the hard X-ray tail include a compact jet, a hot corona and an accretion disc, all of which have been observed in HMXBs \citep{hartog_06, wang_11}. Long term X-ray observations are necessary to confirm the origin of the hard X-ray emission. We have also performed one-zone modelling of the multiwavelength data of HESS J1828-099 and we have successfully reconciled radio, X-ray and TeV data. Although our one-zone model strongly suggests that HESS J1828-099 is a TeV HMGB, the GeV data could not be explained by IC emission using this model. Emission from SNRs associated with molecular clouds, contamination from hitherto undetected nearby pulsar are some of the other possible scenarios that can explain the GeV emission. Nevertheless, based on positional coincidence  and spectral information, as well as the agreeable fit of our one-zone model to the observed multi-wavelength data and the consistency of the best-fit model parameters to that of previously studied HMGBs, we conclude that HESS J1828-099 is the TeV counterpart of the HMXB, thus contributing to the increasing number of TeV HMGBs detected.  Further deep observations in different wavelengths and detailed modelling of the source are needed to confirm the nature of HESS J1828-099.

\begin{acknowledgments}
The authors thank the anonymous reviewer for constructive suggestions regarding the manuscript. This research has made use of archival data (from NuSTAR telescope) and software/tools provided by NASA’s High Energy Astrophysics Science Archive Research Center (HEASARC), which is a service of the Astrophysics Science Division at NASA/GSFC. This work has also made use of public Fermi-LAT data obtained from Fermi Science Support Center (FSSC), provided by NASA Goddard Space Flight Center. A.D.S. thanks Prof. Diego F. Torres and Dr. Vikram Rana for insightful discussions. A.D.S. thanks Partha Pratim Basumallick for help regarding Fermipy data analysis and Hemanth M. for help regarding X-Ray data analysis. N.R. acknowledges Prof. Banibrata Mukhopadhyay for useful discussions. 
\end{acknowledgments}

\software{Fermitools (\url{https://fermi.gsfc.nasa.gov/ssc/data/analysis/scitools/}),
Fermipy (\url{https://fermipy.readthedocs.io/en/latest/}),                    
GAMERA (\url{https://github.com/libgamera/GAMERA}), 
HEAsoft (\url{https://heasarc.gsfc.nasa.gov/docs/software/heasoft/}),
XSPEC (\url{https://heasarc.gsfc.nasa.gov/xanadu/xspec/}),
XSELECT (\url{https://heasarc.gsfc.nasa.gov/ftools/xselect/}),
AstroML (\url{http://www.astroml.org/})}.

\appendix
\section{Monte Carlo Simulations}\label{append:A}

As pointed out in \cite{prota02}, the F-test in some cases does not (even asymptotically) adhere to their nominal $\chi^2$ and F-distributions in many statistical tests common in astrophysics. Thus in this case, the significance of the additional, sub-dominant power law component depicting shock, has been assessed through Monte Carlo simulation method. \texttt{XSPEC} tool \texttt{simftest} was used to perform this task. We used the model depicting accretion component as our \textit{null hypothesis}. The model, which includes the additional power law component with the accretion component, was used as the \textit{alternate hypothesis}. We simulated 1000 trials using \texttt{simftest}, and calculated the change in $\chi^2$ values for the \textit{null hypothesis} and \textit{alternate hypothesis} models. The maximum change in $\chi^2$ ($\Delta\chi^2$) obtained from our simulations is 12.89. The probability of finding the observed change in $\chi^2$ ($\Delta\chi_{obs}^2$ = 9.04) by chance is 6 $\times$ 10$^{-3}$, which corresponds to 4$\sigma$ significance. These results justify the addition of a sub-dominant power law component, which in turn hints towards the presence of shock in the source region of XTE J1829-098. 

\begin{figure}
    \centering
    \includegraphics{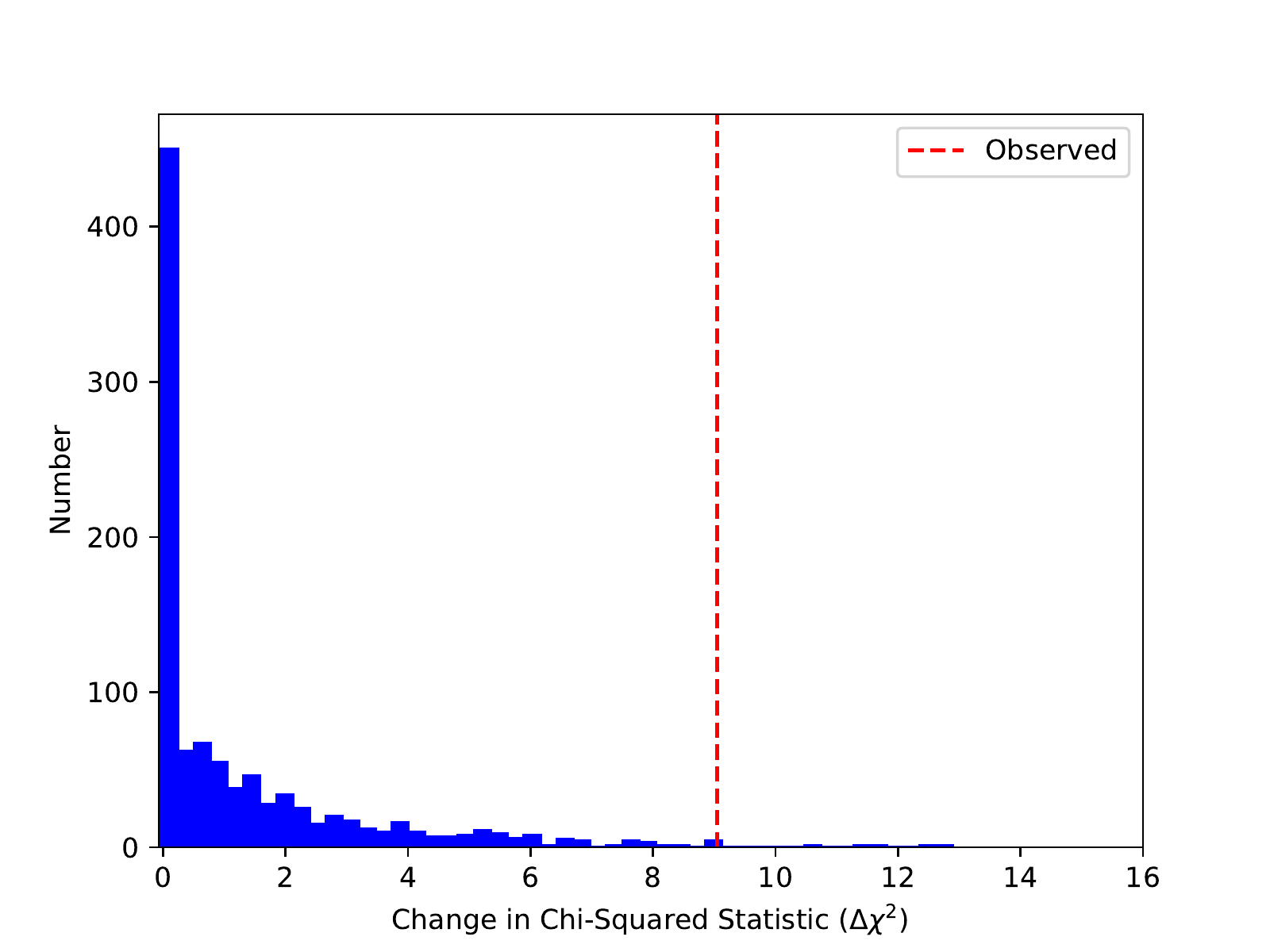}
    \caption{\label{fig4} Results of 1000 Monte Carlo simulations to test the significance of the sub-dominant power law component depicting shock. The blue solid histogram shows the frequency (y-axis) of $\Delta\chi^2$ values (x-axis) obtained in the simulations. The red dashed line shows the observed $\Delta\chi_{obs}^2$ = 9.04.}
    
\end{figure}

\section{\textit{Fermi}-LAT data analysis}\label{append:B}

We have used Fermipy version 0.20.0\footnote{\url{https://fermipy.readthedocs.io/en/latest/}} \citep{wood_17} to reduce and analyze $\sim$ 12 years of PASS 8 LAT data in the energy range of 0.3-500 GeV. Events with zenith angles greater than 90$^\circ$ were excluded from the analysis, to avoid the contamination from Earth's albedo $\gamma$-rays. The instrument response function, Galactic diffuse emission template (galdiff) and isotropic diffuse emission template (isodiff) used in this work were ``P8R3$\_$SOURCE$\_$V2", ``gll$\_$iem$\_$v07.fits" and ``iso$\_$P8R3$\_$SOURCE$\_$V2$\_$v1.txt", respectively. We have used the latest 4FGL catalog \citep{abdol_20} to search for the possible GeV counterpart of HESS J1828-099.

We have extracted the data from the \textit{Fermi}-LAT website\footnote{\url{https://fermi.gsfc.nasa.gov/ssc/data/access/lat/}}, considering a circular region of interest (ROI), having a radius of 10$^\circ$, with the center of the ROI placed at the position of the H.E.S.S. source. Galdiff, isodiff as well as all of the 4FGL sources within a rectangular region of 10$^\circ$ $\times$ 10$^\circ$, centered on HESS J1828-099, were included in the analysis. Pulsar J1828-1007 is within 1$^\circ$ of  the H.E.S.S. source, but it being a radio pulsar \citep{prinz_15}, does not affect our analysis. While analyzing the data, we have kept the parameters of all the 4FGL sources within 4$^\circ$ of the H.E.S.S. source free, including that of galdiff and isodiff. Using the source finding algorithm of Fermipy, we also tried to find point sources around the H.E.S.S. source that are not included in the 4FGL catalog, having a minimum TS value of 25 and minimum separation of 0.3$^\circ$ between any two point sources. However, no plausible point sources in the GeV range were found in the vicinity of the H.E.S.S. source. All the best-fit values of the spatial and spectral parameters of the 4FGL sources, as well as galdiff and isodiff, were determined using maximum-likelihood analysis. Apart from the possible GeV counterpart 4FGL J1830.2-1005, rest of the 4FGL sources, including galdiff and isodiff were considered as background and subsequently subtracted during the analysis. 

\section{Periodicity search}\label{append:C}

Since orbital periodicity is a distinguishable feature of HMGBs, in this work, we searched for periodicity in the $\sim$ 12 years of \textit{Fermi}-LAT $\gamma$-ray data observed from the source 4FGL J1830.2-1005. As discussed in section \ref{sec:intro}, XTE J1829-098 has a possible orbital period of 246 days, as determined from the interval between consecutive outbursts. Since the 4FGL source is the possible GeV counterpart of the HMXB XTE J1829-098, we tried to find similar periodic variation in the light curve of the 4FGL source. To that end, we have produced light curves using the likelihood analysis for time bins of sizes $\approx$ 127 days, balancing low photon statistics and the idea to probe the periodicity of 246 days observed for the XTE source. The background model is considered to be same as that used in Appendix \ref{append:B}. No significant changes in the flux or spectral index were seen in different time bins. A 82.3 days and a 177.7 days binned light curves were also produced, and again, no strong variablity was found in either of the light curves, similar to the previous case.

Next, we searched for periodicity in the 127 days binned light curve, using a generalized Lomb-Scargle algorithm \citep{lomb, scargle}. AstroML package \citep{astroML, astroMLText} was used to search for periodicity in the light curve between 1 to 300 days. We applied the bootstrapping statistical method to calculate the significance levels. 1$\%$ and 5$\%$ significance levels for the highest peak were calculated, determined by 10$^{5}$ bootstrap resamplings. No significant peak confirming any hint of periodicity, was found in the generated power spectra. Bootstrapping indicates that no periodic signal was detected at 1$\%$ or 5$\%$ significance. The same method was reapplied for 82.3 days and 177.7 days binned light curves, but even in those cases, no significant periodicity was detected. The non-detection of periodicity could be either due to inadequate statistics or due to a specific geometrical shape of the binary system that would not produce modulated emission in $\gamma$-rays \citep{hess15}. This is similar to the case of HMGB candidate HESS J1832–093, in which significant periodicity was also not confirmed \citep{Tam_20}. However, a detailed epoch-folding method \citep{Mart_2020} can prove beneficial for finding any periodicity associated with 4FGL J1830.2-1005.

\section{Radio data analysis}\label{append:D}

THOR provides the radio continuum image of $\sim 132$ square degree of the Galactic plane observed with the Karl G. Jansky Very Large Array (VLA) in C array configuration. Out of the eight SPWs covering $1-2$ GHz, two are discarded due to excessive RFI. The other six SPWs (with 128 MHz bandwidth, centred at 1.06, 1.31, 1.44, 1.69, 1.82 and 1.95 GHz) are used to make the continuum images. \citet{thor2} used BLOBCAT \citep{blobcat} to identify sources and extract flux densities, as well as to estimate spectral index values from images at a common resolution of $25\arcsec$. The RMS noise values for individual SPW images are in the range $0.3 - 1.0$ mJy/beam. All the images, the flux density and the spectral index values are available publicly through the latest data release \citep{thor2}; we have used THOR individual SPW images and the combined THOR (VLA C array) and VGPS \citep[VLA Galactic Plane Survey which is VLA D array and Effelsberg 100-m single dish data combined;][]{vgps} image to identify the potential counterpart and adopt the flux density values from THOR catalogue. 

The C band GLOSTAR survey, similarly, covers $\sim 145$ square degree of the Galactic plane observed with the VLA B and D configuration along with the Effelsberg 100-m data to provide zero-spacing information. We use the GLOSTAR survey images from the VLA D configuration, with $18\arcsec$ angular resolution and at an effective frequency of 5.8 GHz (shown in Figure \ref{fig2}). The continuum observations with the VLA were carried out using 16 SPWs with 128 MHz bandwidth each. The data are used to make 8 continuum subimages covering 4.2 - 5.2 GHz and 6.4 - 7.4 GHz. We note that four radio sources are detected within the H.E.S.S. positional error in both THOR and the GLOSTAR survey, but no X-ray counterparts are detected for any of these sources; so it is unlikely that these sources are associated with the H.E.S.S. source.

\bibliography{sample631}{}
\bibliographystyle{aasjournal}

\end{document}